\documentstyle[preprint,aps]{revtex}
\input{psfig}

\begin{document}
%

\draft

\title{Measurements of the reactions $^{12}\rm{C}(\nu_e,e^-)^{12}\rm{N}_{g.s.}$ and 
$^{12}\rm{C}(\nu_e,e^-)^{12}\rm{N}^*$} 

\author{C. Athanassopoulos$^{12}$, L. B. Auerbach$^{12}$,
R. L. Burman$^7$,\\
I. Cohen$^6$, D. O. Caldwell$^3$, B. D. Dieterle$^{10}$, J. B. Donahue$^7$,
A. M. Eisner$^4$,\\ A. Fazely$^{11}$,
F. J. Federspiel$^7$, G. T. Garvey$^7$, M. Gray$^3$, R. M. Gunasingha$^8$,\\
R. Imlay$^8$, K. Johnston$^{9}$,
H. J. Kim$^8$, W. C. Louis$^7$, R. Majkic$^{12}$,
J. Margulies$^{12}$,\\ K. McIlhany$^{1}$, W. Metcalf$^8$, G. B. Mills$^7$,
R. A. Reeder$^{10}$, V. Sandberg$^7$, D. Smith$^5$,\\
I. Stancu$^{1}$, W. Strossman$^{1}$, R. Tayloe$^7$, G. J. VanDalen$^{1}$,
W. Vernon$^{2,4}$, N. Wadia$^8$,\\ 
J. Waltz$^5$, Y-X. Wang$^4$, D. H. White$^7$, 
D. Works$^{12}$, Y. Xiao$^{12}$, S. Yellin$^3$ \\
LSND Collaboration}
\address{$^{1}$University of California, Riverside, CA 92521}
\address{$^{2}$University of California, San Diego, CA 92093}
\address{$^3$University of California, Santa Barbara, CA 93106}
\address{$^4$University of California
Intercampus Institute for Research at Particle Accelerators,
Stanford, CA 94309}
\address{$^{5}$Embry Riddle Aeronautical University, Prescott, AZ 86301}
\address{$^6$Linfield College, McMinnville, OR 97128}
\address{$^7$Los Alamos National Laboratory, Los Alamos, NM 87545}
\address{$^8$Louisiana State University, Baton Rouge, LA 70803}
\address{$^{9}$Louisiana Tech University, Ruston, LA 71272}
\address{$^{10}$University of New Mexico, Albuquerque, NM 87131}
\address{$^{11}$Southern University, Baton Rouge, LA 70813}
\address{$^{12}$Temple University, Philadelphia, PA 19122}
 
\date{\today}
\maketitle
\begin{abstract}
Charged current reactions of $\nu_e$ on $^{12}\rm{C}$ have been studied using a 
$\mu^+$ decay-at-rest $\nu_e$ beam from the Los Alamos Meson Physics Facility.
More than 500 events from the exclusive reaction
$^{12}\rm{C}(\nu_e,e^-)^{12}\rm{N}_{g.s.}$ were measured in a large Liquid Scintillator Neutrino
Detector (LSND). The observed energy dependence of the cross section and the
angular distribution of the outgoing electron agree well with theoretical 
expectations.  Measurements are also presented for inclusive transitions to $^{12}\rm{N}$
excited states,  $^{12}\rm{C}(\nu_e,e^-)^{12}\rm{N}^*$ and compared with
theoretical expectations.
Results are consistent with a recent 
Continuum Random Phase Approximation (CRPA) calculation.

\end{abstract}

\section{Introduction}

There are two principal reasons for measuring low energy (E$_\nu < $52 MeV)
neutrino-nucleus
scattering. First the extracted cross section provides unique insight into 
nuclear dynamics.   The yield depends on dynamics as expressed by nuclear 
axial-vector and vector currents, thereby providing additional 
information beyond
the vector currents which are obtained from electron-nucleus scattering.
Next, the neutrino-nucleus cross sections are required both for calculating 
certain astrophysical processes and for characterizing the response
of neutrino detectors.   In the former case, neutrino-nucleus interactions in 
the outer shells of stars undergoing supernova explosions cause further
nucleosynthesis to occur~\cite{Woosley}.
Neutrinos of all flavors are produced in the interior of the star during 
supernova collapse.  These neutrinos undergo interactions with
nuclei changing the nuclear composition of 
the star via charge-changing processes and the excitation of particle unstable
nuclear states. In the latter case, many of present day active neutrino 
detectors are composed of $^{12}$C or $^{16}$O nuclei, in addition to 
$^{1}$H  or $^{2}$H, and thus require accurate knowledge of these 
neutrino-nucleus cross sections to reliably interpret the detector output.

At the present time, relatively few measurements of neutrino-nucleus cross 
sections exist.   Neutrino-carbon cross sections for neutrinos created
from the decay of stopped positive muons have been measured only twice before
this experiment.  E225~\cite{Krakauer} 
at LAMPF and the KARMEN collaboration~\cite{Bodmann} at ISIS 
facility of the Rutherford Laboratory have measured the cross section
for the exclusive reaction $^{12}\rm{C}(\nu_e,e^-)^{12}\rm{N}_{g.s.}$ and
for the more inclusive reaction $^{12}\rm{C}(\nu_e,e^-)^{12}\rm{N}^*$
to all the other accessible  $^{12}\rm{N}$ final states.
The yield to the $^{12}\rm{N}$ ground state dominates the total yield, 
as it is the only allowed ($\ell=0$) transition that occurs in this process.
In this paper we report on our measurement of the same processes.
Our results are more accurate than the earlier 
measurements~[2-4] 
and in good agreement with them.  We have also measured the
angular distributions of the electrons with respect to the $\nu_e$ 
direction for these processes as well as the energy dependence 
of the ground state
transition.  All are in excellent accord with expectation.

Calculation of neutrino-$^{12}\rm{C}$ cross sections have attracted a good
deal of attention~[5-9].
The cross section for producing the $^{12}\rm{N}$
ground state can be calculated  to an accuracy of 
2\% as it can be represented in terms of form factors~\cite{Fukugita} that
can be reliably extracted from other measurements.  Calculating the
inclusive yield to the excited states is a far less certain procedure.
The Fermi Gas Model (FGM) is not applicable as the momentum transfers 
(Q $<$ 100 MeV/c) are much smaller than the Fermi momentum (200 MeV/c) in
carbon.   Thus it is necessary to employ a model that reliably 
reflects the nuclear dynamics.   The most recent work~\cite{Kolbe} uses a model
that includes the residual particle-hole interaction via the random-phase
approximation (CRPA).  The authors have argued that the CRPA calculation
for this process should be accurate once the parameters have been
determined via fitting to a comparable process such as, say, negative
muon capture on $^{12}\rm{C}$. Their calculations correctly
represent the experimental results we present in this paper
but it should be mentioned that their prediction~\cite{Kolbe} 
for our preliminary results~\cite{Albert,kim} on the inclusive yield from 
$^{12}\rm{C}(\nu_\mu, \mu^-)^{12}\rm{N}^*$ obtained using higher energy 
$\nu_\mu$ from  $\pi^+$ decay in flight is too large by nearly a factor of two.
This discrepancy has generated considerable theoretical 
interest~\cite{Umino,Kolbe2} but
remains unexplained.  We will publish our final results on this measurement
in a subsequent paper~\cite{chris}.

Finally, we note the relevance of the analysis presented in this paper to
the evidence for neutrino oscillations previously presented by the LSND
collaboration~\cite{Bigpaper2}. 
The excellent agreement found with expectations and with
previous experiments for the reaction 
$^{12}\rm{C}(\nu_e,e^-)^{12}\rm{N}_{g.s.}$ provides important
confirmation of our understanding of the neutrino beam and the detector
performance. The neutron analysis presented in section VII likewise provides
a valuable test of the techniques used for neutron identification.

\section{The Neutrino Source}

The data reported here were obtained in 1994 and 1995 at the Los Alamos Meson
Physics Facility (LAMPF) using neutrinos produced at the A6 proton beam stop.
The neutrino source is described in detail elsewhere~\cite{Bigpaper1}. 
This facility is now the Los Alamos Neutron Science Center (LANSCE).
The beam stop consists of a 30 cm water target surrounded by steel shielding
and followed by a copper beam dump.  The high-intensity 800 MeV proton beam
from the linear accelerator generates a large pion flux from the water target.
The flux of $\nu_e$ used for the measurements reported here arise from the
decay at rest (DAR) of stopped $\pi^+$ and $\mu^+$.  This decay chain yields almost 
equal intensities of $\nu_e, \bar{\nu}_\mu$ and $\nu_\mu$ with the well determined
energy spectra shown in Fig.~1.
  
The corresponding decay chain for $\pi^-$ and 
$\mu^-$ is highly suppressed due to three factors.  First, production of $\pi^-$
is approximately eight times smaller than for $\pi^+$ at LAMPF.  Second,
$\pi^-$ which stop are absorbed by nuclear interactions.  Finally, most $\mu^-$ 
which stop are absorbed before they can decay. These stopped $\mu^-$ arise from
$\pi^-$ which decay in flight since $\pi^-$ which stop are absorbed.
Approximately 3.4\% of the $\pi^+$ and 5\% of the $\pi^-$ decay in flight (DIF)
yielding fluxes of $\nu_\mu$ and $\bar{\nu_\mu}$ that are used for the complementary
$\nu_\mu C$ measurements described elsewhere~\cite{kim,chris}.

The LAMPF beam dump has been used as the neutrino source for previous experiments E31~\cite{E31},
E225~\cite{Krakauer} and E645~\cite{Freedman}.   A calibration experiment, E866~\cite{Allen}, 
measured the rate of stopped $\mu^+$ from  a low intensity proton beam incident on 
an instrumented beam stop.
The rate of stopped $\mu^+$ per incident proton was measured as a function of 
several variables and used to fine tune a beam dump simulation program~\cite{Burman}.  This 
simulation program can then be used to calculate the flux for any particular beam
dump configuration. The calibration experiment determined  the DAR
flux to $\pm 7$\% for the proton energies and beam stop configurations used at LAMPF. 
This 7\% uncertainty  provides the largest
source of systematic error for the cross sections presented here.  
It is worth mentioning that the measurements of the $\nu_e \rm{C}$ cross section 
made by the KARMEN collaboration~\cite{Bodmann} and E225~\cite{Krakauer} also rely on this
calibration experiment and beam dump simulation program.
The LAMPF proton beam typically had a current of 800  $\mu$A and 
an energy of approximately 770 MeV at the A6 beam stop. 
The integrated beam current was 5904 C in 1994 and 7081 C in 1995.  The calculated 
ratio of stopped $\mu^+$ per proton was 0.090 and 0.084 for 1994 and 1995 respectively
with the lower ratio for 1995 arising because the water target was out for 32\% of the
1995 data.  Upstream targets A1 and A2 contribute 1.4\% to the DAR flux.  The DAR $\nu_e$
flux averaged over the LSND detector was then $3.04 \times 10^{13} \rm{cm}^{-2}$ for 
1994 and $3.43 \times 10^{13} \rm{cm}^{-2} $ for 1995.

\section{The LSND Detector}

The detector is located 29.8 m downstream of the proton beam stop at an angle 
of $12^\circ$ to the proton beam.  Fig.~2
shows a side-view of the setup.   Approximately 2000 $\rm{g/cm^2}$ of shielding above
the detector attenuates the hadronic component of cosmic rays to a negligible level.
Enclosing the detector, except on the bottom, is a highly efficient liquid 
scintillator veto shield which is essential to reduce contributions from 
the cosmic ray muon
background to a low level.  The detector is also well shielded from the beam 
stop so that beam associated 
neutrons are attenuated to a negligible level.  Ref.~\cite{Bigpaper1} 
provides a detailed description of the detector, veto and data acquisition
system which we briefly review here.

The detector is a roughly cylindrical tank containing 167 tons of liquid scintillator and
viewed by 1220 uniformly spaced 8'' Hamamatsu PMTs 
covering $\sim$25\% of the surface inside the tank wall.
When the deposited
energy in the tank exceeds a threshold of approximately 4 MeV electron-equivalent
energy and there are fewer than 4 PMT hits in the veto shield, then
the digitized time and pulse height of
each of these PMTs (and of each of the 292 veto shield PMTs) are recorded. 
  A veto, imposed for 15.2 $\mu$s following
the firing of $>$ 5 veto PMTs, substantially reduces ($10^{-3}$) 
the large number of background events arising from 
the decay of cosmic ray muons that stop in the detector.  
Activity in the detector or veto shield during the 
51.2 $\rm{\mu s}$ preceding a primary trigger is also recorded provided there are $>$17 detector 
PMT hits or $>$5 veto PMT hits.  This activity information is used in the analysis to
further reject events arising from muon decay.  Data after the primary event are 
recorded for 1 ms with a threshold of 21 PMTs (approximately 0.7 MeV equivalent). 
This low threshold is necessary for neutron identification as described below.
The detector operates without reference to the beam spill, but 
the state of the beam is 
recorded with the event. Approximately 93\%
of the data is taken between beam spills. This allows an accurate
measurement and subtraction of cosmic ray background surviving the event
selection criteria.

The detector scintillator consists of mineral oil ($\rm{CH}_2$) in which is 
dissolved a small concentration (0.031 g/l) of b-PBD~\cite{Reeder}.
This mixture allows the separation of $\check{\rm C}$erenkov light and scintillation light
and produces about 33 photoelectrons per MeV of electron energy deposited
in the oil.
The combination of the two sources of light provides direction information
and makes particle identification (PID)
possible for relativistic particles.
Identification of neutrons is accomplished through the detection of 
the $2.2\,{\rm {MeV}}$ 
$\gamma$ from neutron capture on free protons. Note that the oil consists
almost entirely of carbon and hydrogen. Isotopically the carbon is 1.1\% 
$^{13}$C and 98.9\% $^{12}$C.

The veto shield encloses the detector on all sides except the bottom.
Additional counters were placed below the veto shield 
after the 1993 run to reduce cosmic 
ray background entering through the bottom support structure.
These counters around the bottom support structure are referred to as bottom counters.
The main veto shield \cite{veto} consists of a 15-cm layer of liquid 
scintillator in an external tank and 15 cm of lead shot in an internal 
tank.
This combination of active and passive shielding
tags cosmic ray muons that stop in the lead shot.
A veto inefficiency $<10^{-5}$ is achieved with this detector
for incident charged particles. The veto inefficiency is large 
for incident cosmic ray neutrons.

\section{Analysis Techniques}

Each event is reconstructed using the hit time and pulse height of all hit
PMTs in the detector~\cite{Bigpaper1}. The present analysis  relies on the 
reconstructed energy, position, direction 
and particle ID parameter, $\chi_{tot}$.
  The particle direction is determined from the Cerenkov cone.  The parameter
$\chi_{tot}$ is used to distinguish electron events from events arising from 
interactions of cosmic ray neutrons in the detector.  Fortunately, it is possible
to measure the response of the detector to electrons and neutrons in the energy 
range of interest for this analysis. We also make use of a detailed Monte Carlo
simulation, LSNDMC~\cite{mc}, which was written to simulate events in the detector using
GEANT. 

The response of the detector to electrons was determined from a large, essentially
pure sample of electrons (and positrons) from the decay of stopped cosmic 
ray $\mu^\pm$ in the detector.  The known energy spectra for electrons from
muon decay was used to determine the absolute energy calibration including its 
small variation over the volume of the detector.  The energy resolution was
determined from the shape of the electron energy spectrum 
 and was found to be
6.6\% at the 52.8 MeV endpoint. The position and direction resolution obtained from
the LSNDMC simulation are approximately 30 cm and $17^\circ$ respectively for 
electrons in the energy region of interest, 16 to 35 MeV.  The precision of 
position reconstruction has been checked from a comparison of the reconstructed 
positions of the $\mu^-$ and the decay $e^-$ of a large sample of 
$\nu_\mu \rm{C} \rightarrow \mu^- \rm{X}$ events~\cite{Bigpaper1}. 
The accuracy of the direction
measurement is discussed more in section VI.  

There are no tracking devices in the LSND detector and thus event positions must be determined
solely from the PMT information.  The reconstruction process determines an event position
by minimizing a function $\chi_r$ which is based on the time of each PMT hit corrected for
the travel time of light from the assumed event position to the 
PMT~\cite{Bigpaper1}.
This reconstruction procedure was found to systematically shift event positions away
from the center of the detector and thus effectively reduces the fiducial volume~\cite{Bigpaper2}.
In the analysis presented in this paper a fiducial cut is imposed by requiring D $>$ 35 cm,
where D is the distance between the reconstructed event position and the surface
tangent to the faces of the PMTs.  

The effect of the reconstruction bias on the fiducial acceptance
 was determined from the analysis of a
sample of stopping muon events for which both the muon and the subsequent
decay electron were detected.  No fiducial cut was imposed on either the
muon or the electron so that essentially all muons which stopped in the
scintillator and decayed were included.  For comparison a sample of
simulated stopping muon events was generated using LSNDMC.  The observed
and generated distributions of the distance D were compared for electrons satisfying a 
minimum energy requirement.  The observed distribution was found to be 
shifted
outward relative to the generated distribution.  Five independent analysis
of this type yielded the acceptance factor of $0.85 \pm 0.05$ for D$>$35 cm due
to the reconstruction bias.  There is independent support for this 
conclusion.
A new reconstruction procedure has been developed which relies both on
PMT pulse height and timing information, and is  expected to
be less biased.  
   This new reconstruction procedure calculates the likelihood for 
   the observed PMT charge distribution and time distribution as a function 
   of position.  
   The final position is then determined by maximizing the likelihood.
Comparison of positions obtained with the new and the standard
reconstruction procedures indicate a pushing out effect in good agreement with
that obtained from the stopping muon analysis.

The particle identification procedure is designed to separate particles with
velocities well above Cerenkov threshold from particles below Cerenkov threshold by
making use of the four parameters defined in Ref.~\cite{Bigpaper1}.  Briefly, $\chi_r$ and
$\chi_a$ are the quantities minimized for the{ determination of the event position
and direction, $\chi_t$ is the fraction of PMT hits that occur more than 12 ns
after the fitted event time and $\chi_{tot}$ is proportional to the product of 
$\chi_r, \chi_a$ and $\chi_t$.  For the present analysis we use only $\chi_{tot}$.
Fig.~3 shows the $\chi_{tot}$ distributions for 
electrons from stopping $\mu$ 
decay and for cosmic ray neutrons with electron equivalent energies in the $16 < \rm{E_e} < 35$ MeV range.  
For a neutron $\rm{E_e}$ is the equivalent electron energy corresponding to the observed total
charge.   In the present analysis we eliminate most cosmic
ray neutron background by requiring $\chi_{tot}<$ 0.85.

The presence of a neutron can be determined by the neutron capture 
reaction $\rm{n + p \rightarrow d + \gamma}$.
The mean capture time in the LSND detector is expected to be 186 $\rm{\mu s}$, essentially
independent of the initial neutron energy.
Three variables are used to identify a capture $\gamma$ correlated with a neutron 
in the primary
event: the number of PMT hits for the $\gamma$, the distance of the $\gamma$ from
the primary event and the time of the $\gamma$ from the primary event.  
Fig.~4 shows the distributions of these 
variables for correlated $\gamma$s and{
for uncorrelated (accidental) $\gamma$s. A likelihood technique, discussed in 
Ref.~\cite{Bigpaper2}, has been developed to separate the correlated component due to
neutrons from the uncorrelated component.  An approximate likelihood ratio
 $R \equiv {\cal L}_{cor}/{\cal L}_{uncor}$
is calculated for each event from 
the three measured variables. If there is no $\gamma$ within 1 ms and
2.5 m from the primary event then R$=0$ for the event.
The expected distributions of R are shown in Fig.~5 for a correlated sample
(every event has one neutron) and for an uncorrelated sample (no event has
a neutron).  
     The correlated R distribution was found to be almost independent of
     event position within the fiducial volume~\cite{Bigpaper2}. 
     The accidental gamma rate
     is higher near the bottom front corner of the detector then elsewhere,
     but the shape of the uncorrelated R distribution has little position
     dependence.
In the present paper we use the $\gamma$ analysis solely to verify
that the events in the $^{12}\rm{C}(\nu_e, e^-)^{12}\rm{N}$ samples are not 
accompanied by neutrons and to study the cosmic ray background.
 The measured R distribution is fit to a mixture of the
two distributions shown in Fig.~5 and the fraction of events with neutrons
is obtained.  

Beam-off data taken between beam spills plays a crucial role in the analysis of 
this experiment.  Most  event selection criteria are designed to reduce the cosmic
ray background while retaining high acceptance for the neutrino process of 
interest.  Cosmic ray background which remains after all selection criteria have
been applied is well measured with the beam-off data and subtracted using the 
duty ratio, the ratio of beam-on time to beam-off time.  This ratio
was 0.080 for 1994 and 0.060 for 1995. Beam-on and beam-off data have been 
compared to determine if there are any differences other than those
arising from neutrino interactions.  No differences are found in trigger rates, veto
rates or various accidental rates, including accidental $\gamma$ rates, and
there is no evidence of beam neutrons at any energy.  From these and more
detailed comparisons we find no indication of any non-neutrino induced beam background 
or any problem with the beam-off subtraction procedure.

\section{The transition to the $^{12}\rm{N}$ ground state} 

The reaction $\nu_e +~^{12}\rm{C}\rightarrow ^{12}\rm{N}_{g.s.} + e^-$ is identified 
by the detection of the $e^-$ followed, within 45 ms, by the positron from the 
$\beta$ decay of the $^{12}\rm{N}_{g.s.}$. Transitions to excited states of $^{12}$N decay by prompt
proton emission and thus do not feed down to the $^{12}$N ground state or contribute to the
delayed coincidence rate.
The scattered electron has a maximum kinetic energy of 
35.5 MeV due to the Q value of 17.33 MeV.  The beta decay has a mean lifetime of 15.9 ms and
 maximum positron kinetic energy of 16.33 MeV.  The cross section to the $^{12}\rm{N}$
ground state has been calculated by several groups~[5-9].
The form factors required to calculate the cross section are well known from a variety
of previous measurements.
This cross section and the known $\nu_e$ flux are used to obtain the expected electron
kinetic energy spectru.   Fig.~6
shows the observed electron energy distribution in the beam-on, beam-off and 
beam-excess samples for events with an identified beta decay.
Fig.~6(c)
compares the expected and observed energy distributions.

The selection criteria and corresponding efficiencies for the electron 
are shown in Table I.  The reconstructed
electron position is required to be a distance $D > 35$~cm from the surface tangent to the
faces of the PMTs.  There are $3.65 \times 10^{30}$ $^{12}$C nuclei within this fiducial volume.
A lower limit on the electron energy of 16.0 MeV eliminates
the large cosmic ray background from $^{12}\rm{B}$ beta decay  as well
as most 15.1 MeV gamma rays from the neutral current excitation of carbon.   
The $^{12}\rm{B}$ nuclei arise from the absorption of stopped $\mu^-$ on $^{12}\rm{C}$ nuclei in
the detector.

The
past activity cut is designed to reject most electron events arising from
cosmic ray muons which stop in the detector and decay.  This background has a time 
dependence given by the 2.2 $\rm{\mu s}$ muon lifetime.
The past activity selection criteria reject all events with activity within the past 
20 $\rm{\mu s}$. Events with activity between 20 $\rm{\mu s}$ and 35 $\rm{\mu s}$ in the 
past are rejected if for the activity either (a) the detector charge is greater than
3000 photoelectrons ($\sim 100$ MeV) or (b) the number of tank hits is greater than 100.  Beyond 
35 $\rm{\mu s}$ no cut is applied.     
Fig.~7 shows the distribution of time to the closest past activity for beam-off
and beam-excess events passing these criteria.
Only events with past activity between 20 and 50 $\rm{\mu s}$ are plotted.
The relatively loose cut applied in the 20 to
35 $\rm{\mu s}$ time interval is still adequate to reject most muons surviving this long.
This can be seen in Fig.~7(a) where 
the rate of beam-off events is comparable above
and below 35 $\mu$s and there is no indication of a time dependence corresponding to the 
2.2 $\rm{\mu s}$ muon lifetime.
The rate of beam-excess events, shown in Fig.~7(b), 
is also
comparable above and below 35 $\rm{\mu s}$, consistent with the small calculated loss of acceptance (4.6\%)
for the 20 to 35 $\mu$s interval. 
In-time veto cuts are designed to reject events arising directly from cosmic ray particles which enter the detector.  Events with more than 3 veto hits or any bottom counter coincidence during the 500 ns event window are eliminated .  The beam-off subtraction procedure removes
the small cosmic ray background which survives the above cuts.

The acceptances for the election selection criteria are shown in Table I separately
for 1994 and 1995 data samples.  
  The acceptances for the 
past activity and in-time veto cuts are obtained by applying these cuts to a large
sample of events triggered with the laser used for detector calibration.
These laser events are spread uniformly through the run and thus average over the small
variation in run conditions.  The acceptance for the 15.1 $\rm{\mu s}$ 
trigger veto is included in
the past activity efficiency. 
A sample of Michel electrons (electrons from the decay of stopped $\mu^{\pm}$)
 was analyzed to obtain the acceptance of electrons 
for the $\chi_{tot}$ particle identification cut.  The Michel electrons were given weights as
a function of energy so that the weighted spectrum agrees with the observed 
energy spectrum for the $\nu_e \rm{C}$ sample.  Similarly, events are weighted as a function of position.  
Fig.~8 compares the $\chi_{tot}$ distribution of the electrons in the $\nu_e \rm{C}$ sample 
with the weighted Michel sample.  The agreement is excellent.   

Table II gives the 
selection criteria and efficiencies for the $^{12}\rm{N}$ beta decay positron.
Fig.~9 shows the observed beta
decay time distribution compared with  the expected 15.9 ms lifetime.
Fig.~10 shows the distance between the reconstructed electron and positron
positions for 
the beam-excess sample.  A cut was applied at 100 cm resulting in an acceptance of
$(96 \pm 2)\%$. Following an electron produced 
by a neutrino interaction an uncorrelated particle, such as the positron from
$^{12}B$ beta decay, will occasionally satisfy all the positron criteria including
the requirements of time (45 ms) and spatial (1 m) correlation with the
electron.   The probability of such an accidental coincidence can be precisely measured from the Michel electron sample.
 The background from this source is also shown in Fig.~9 and 10.
The efficiency of 81.5\% caused by the 
15.1 $\mu$s veto and the trigger dead time of 3\% are the same as for the electron.
 Positrons with 4 or more in-time veto hits or any bottom veto coincidence 
are rejected.  Fig.~11 shows the observed $\chi_{tot}$ distribution for the
positron for the beam-excess sample.   No cut on $\chi_{tot}$ is applied.
The energy distribution of the positron is calculated 
from the $^{12}N$ beta decay using

\begin{equation}
\rm{{dN \over dE_{e}}= P_{e}E_{e}(E_{max}-E_{e})^{2}\times {2\pi\eta \over (\
e^{2\pi\eta}-1)}}
\end{equation}
\noindent
where $\eta={Z\alpha \over \beta_{e}}$ and $\rm{E_e}$ is the total positron energy (including
rest energy). The $^{12}\rm{N}$ decays to the ground state ($\rm{E_{max}} = 16.84$~MeV) 94.6\% of the time. 
Beta decay transitions to the excited states of carbon 
are 1.9\% ($\rm{E_{max}} = 12.38$~MeV, followed by
a 4.4 MeV $\gamma$),  2.7\% ($\rm{E_{max}}$ = 9.17 MeV)  and 
0.8\% ($\rm{E_{max}}$ = 6.5 MeV)~\cite{beta}. 
The positron annihilates with an electron after stopping.
The Monte Carlo was used to generated expected distributions for the positron energy and
for number of hit PMTs.  There was a trigger requirement of 100 PMT hits for 1994 and 
75 PMT hits for 1995.
Fig.~12 compares the observed and expected positron 
energy distributions. The good agreement shows that the energy calibration
is valid for these low energy electrons.

Table III provides a breakdown of the number of events satisfying the selection criteria
as well as the acceptances, the neutrino flux and the resulting flux averaged 
cross section for both years of data.  For the complete data sample the flux averaged cross 
section is 

$<\sigma> = (9.1 \pm 0.4 \pm 0.9)\times 10^{-42}~\rm{cm}^2$ 

\noindent
where the first error is statistical and the second is systematic.  The two dominant 
sources of systematic error are the neutrino flux (7\%) discussed in section II and the 
effective fiducial volume (6\%) discussed in section IV. 
The spatial distribution of the electrons is shown in Fig.~13.
   The measured
cross section decreases by $(2.7 \pm 2.2)$\% when the fiducial volume is reduced by
requiring that the electron be at least 50 cm (instead 35 cm) from the surface
of the PMT faces.
For comparison the two previous measurements, the LSND result 
and several theoretical predictions
for the flux averaged cross section are presented in Table IV.  
They are all in agreement with each other.

For this reaction to the $^{12}\rm{N}$ ground state it is also
straightforward to measure the energy dependence of the cross section.  The 
recoil energy of the $^{12}\rm{N}$ nucleus is negligible and thus $\rm{E_\nu = E_e + 17.3~MeV}$
where $E_e$ is the electron kinetic energy.  Fig.~14 shows that the
measured cross section agrees with the requisite energy dependence~\cite{Fukugita}.
The expected shape shown in the figure includes the effects of detector
resolution and acceptance obtained from the Monte Carlo.
Fig.~15 shows the observed and expected~\cite{Fukugita,Akolbe} 
angular distribution between the electron and the incident neutrino. 
The only previous measurement~\cite{Allen2} had very limited angular
acceptance.    

\section{Transitions to excited states of $^{12}\rm{N}$}

Electrons below 52 MeV are expected to arise from four major neutrino 
processes: $^{12}\rm{C}(\nu_e, e^-)^{12}\rm{N}_{g.s.}$, 
$^{12}\rm{C}(\nu_e, e^-)^{12}\rm{N}^*$, 
$^{13}\rm{C}(\nu_e, e^-)^{13}\rm{X}$ and neutrino electron elastic scattering.
The expected energy and angular distributions of these processes are shown in
Fig.~16 and Fig.~17, respectively.  
The different event characteristics of these reactions are 
used to select a sample due primarily to the reaction $^{12}\rm{C}(\nu_e, e^-)^{12}\rm{N}^*$.
This sample is then used to determine the flux averaged cross section and the electron energy
and angular distributions for this reaction.

All three types of DAR neutrinos ($\nu_e, \nu_\mu$ and $\bar{\nu}_\mu$) elastically scatter off
electrons in the detector but the rate is dominated by $\nu_e e^-$ scattering~\cite{Allen3}.
The contribution due to DIF $\nu_\mu$ and $\bar{\nu}_\mu$ scattering on electrons is negligible.
The scattered electron for this process is strongly forward peaked as 
shown in Fig.~17, and 
thus such events can largely be eliminated with an angle cut.

A second background arises from the interaction of $\nu_e$ on $^{13}$C nuclei (1.1\% of the
carbon).  The expected number of events obtained from the calculated 
cross section~\cite{c13,Donnelly} for this process is fairly small.
The $Q$ value is 2.1 MeV and thus about half of the background can be eliminated by requiring an
electron energy below 34 MeV.  We use the cross section calculated by Kubodera~\cite{c13}
, $0.525 \times 10^{-40} \rm{cm}^2$, and conservatively assign a 50\% uncertainty 
to this number.

The reaction $^{12}\rm{C}(\nu_e, e^-)^{12}\rm{N}_{g.s.}$ is also a source of background since
the $e^+$ from the beta decay of $^{12}\rm{N}_{g.s.}$ is not always identified.
Any event with an identified $e^+$ in delayed coincidence is of course excluded.  
The background of events with unidentified $e^+$ is calculated using the positron acceptance
given in Table II and subtracted.

Slightly tighter selection criteria are needed for the electron in this analysis than
was the case for the $^{12}\rm{N}_{g.s.}$ analysis where the requirement of an $e^+$ in 
delayed coincidence substantially reduced the background. Fig.~18 
shows the 
measured electron energy distribution for beam-off and beam-excess events excluding identified
$^{12}\rm{N}_{g.s.}$ events.  For this analysis the electron energy is required to be
between 20 and 34 MeV, a region that contains 58\% of the expected $^{12}\rm{N}^*$ signal.
The lower limit of 20 MeV excludes 
$^{12}\rm{B}$ beta decay induced by the capture of $\mu^-$ cosmic rays on
$^{12}\rm{C}$ and is 
enough above the 15.1 MeV $\gamma$ from the neutral current excitation of $^{12}\rm{C}$
 that most events from this source are eliminated.  The upper limit of 34 MeV minimizes 
the background from the process  $^{13}\rm{C}(\nu_e, e^-)^{13}\rm{X}$ as well as from the 
possible oscillation signal~\cite{Bigpaper2} seen mostly above this energy. 

A slightly tighter fiducial requirement is also imposed.  Fig.~19
 shows the $y$ distribution for 
beam-off and beam-excess events.  The requirement $y > -120$ cm removes the region where a
large beam-off subtraction results in large statistical errors.  Fig.~20 shows the distribution
of the cosine of the angle between the electron and the incident neutrino.   
The expected 
distribution from all processes is also shown in Fig.~20.
The requirement $\rm{cos~\theta < 0.9}$ removes the forward peak due to $\nu e$ scattering. 
The selection criteria and acceptance for this analysis are shown in Table V.
The total number of beam-on and beam-excess events satisfying these criteria,  
the number of background events,
and the resulting numbers of events and cross section for the process   
$^{12}\rm{C}(\nu_e, e^-)^{12}\rm{N}^*$ are shown in Table VI.  The flux averaged
cross section obtained from the full data sample is 

$<\sigma> = (5.7 \pm 0.6 \pm 0.6)\times 10^{-42}~\rm{cm}^2$. 

\noindent
There are several contributions to the systematic error. The 7\% flux uncertainty and
6\% uncertainty in the effective fiducial volume have been described previously.  There
is a 4\% uncertainty arising from the 50\% error in the $^{13}\rm{C}$ cross section.
The uncertainty in the $e^+$ acceptance for the $^{12}\rm{N}_{g.s.}$ 
background subtraction leads to a 5\% uncertainty in the $^{12}\rm{N}^*$ cross section.
The uncertainty in the duty ratio results in a 2\% error in the cross section.
We also rely on the theory to obtained the fraction of events (58.2\%) with
electrons in the region 20 MeV $< \rm{E}_e <$34 MeV.  
Cascade gammas arising from nuclear transitions contribute to 
the measured energy for some events and slightly increase the acceptance
of the electron energy cut. There is an estimated 3\% uncertainty in the 
cross section from this effect.
An excess of events satisfying criteria for neutrino oscillation
has been reported~\cite{Bigpaper2}. 
Most of this possible oscillation signal is 
above the 34 MeV energy requirement used in this analysis.
The resulting background for the N$^*$ cross section would be 1 to 5\% 
depending on the value of $\delta\rm{m}^2$. 
The spatial distribution of events agrees well with expectations.
The measured cross section decreased  by  $(3.9 \pm 2.8)$\% when the fiducial volume is reduced by
requiring that the electron be at least 50 cm (instead of 35 cm) from the surface formed
by the PMT faces.  Similarly, requiring $ y > -100$ cm (instead of 
$ y > -120$ cm) reduces 
the cross section by $(0.4 \pm 1.8)$\%.  Removing the explicit $y$ cut reduced 
the cross section by $(4.2 \pm 3.7)$\%. 
The flux averaged cross section measured by LSND is compared with other measurements and 
with theoretical calculations in table VII.  The value obtained by LSND agrees well
with a recent CRPA calculation~\cite{Kolbe} and with both earlier 
experimental results within 
errors~\cite{Drexlin,Krakauer}.

The total charged current cross section for $\nu_e$ interactions on 
$^{12}\rm{C}$ can be obtained by combining the measurement presented here on
transitions to $^{12}\rm{N}^*$ with the measurement presented in section V for
the process $^{12}\rm{C}(\nu_e,e^-)^{12}\rm{N}_{g.s.}$. The resulting flux averaged cross section for the process $^{12}\rm{C}(\nu_e,e^-)^{12}\rm{N}$ is 

$<\sigma> = (14.8 \pm 0.7 \pm 1.4)\times 10^{-42}~\rm{cm}^2$. 

\noindent
The errors given take into account the correlations between the two measurements.
The dominant sources of systematic error are the neutrino flux (7\%) and 
the effective fiducial volume (6\%).

There is a clear forward peak due to neutrino electron elastic scattering in Fig.~20.
A measurement of this process will be reported in a future publication after
we take one more year of data.
The good agreement of the observed and expected number of events
in the forward peak indicates that the direction determination is reliable.
For the slowly varying angular distribution for electrons from $\nu_e$ carbon interactions the
angular resolution is more than adequate.
The angular distribution was also measured for a sample of Michel electrons.  Small
systematic distortions ($\sim$10\%) related to the detector geometry were observered in the
angular distribution.  These distortions were well reproduced in the detector Monte Carlo
and are corrected for in the distributions shown. 

The measured and expected distributions of electron energy and
cos~$\theta$ are shown in Fig.~21 and Fig.~22, respectively
for the process  $^{12}\rm{C}(\nu_e, e^-)^{12}\rm{N}^*$. 
The cos~$\theta$ distribution is enhanced in the backward direction as expected~\cite{Donnelly,Akolbe}.
The backward peaking of the angular distribution is largely a result of the negative
parity of the N$^*$ states expected to contribute, $2^-$ levels at 1.20 and
 4.14 MeV and $1^-$ levels at 6.40 and 7.68 MeV.   
The $\ell=1$ angular momentum transfer to the A$=$12 system favors
momentum transfer of approximately 100 MeV/c, and hence the backward peaking.

\section{Neutron analysis}
The electron event samples were also analyzed to determine the fraction of 
events with an associated  neutron. The presence of a neutron 
is determined by detection of the 
$\gamma$ from the reaction $\rm{n + p \rightarrow d + \gamma}$, using the procedure
discussed in section IV.
No neutron production is associated with two of the reactions 
previously discussed,
$^{12}\rm{C}(\nu_e,e^-)^{12}\rm{N}_{g.s.}$ and neutrino electron elastic scattering. Further,
there are kinematic constraints on neutron production by the interaction of 
DAR $\nu_e$ on $^{12}\rm{C}$ and $^{13}\rm{C}$.  Neutron production is not possible
for  $^{12}\rm{C}$ and $^{13}\rm{C}$, respectively, for electron energies above 21 and 31 MeV.
Nevertheless, the neutron analysis provides a useful check on our understanding of 
the sources of inclusive electrons in our data samples.

The distribution of the likelihood ratio R for correlated $\gamma$s from 
neutron capture is very different than for uncorrelated (accidental) 
$\gamma$s as shown in
Fig.~5.  The measured R distribution for 
a data sample can be fit to a mixture of the
two sources of $\gamma$s to determine the fraction of events with a neutron.
Fig.~23 shows the R distribution for the clean sample of
$^{12}\rm{C}(\nu_e,e^-)^{12}\rm{N}_{g.s.}$ beam-excess events, discussed in Section V, 
for which neutron production is not possible.  The best fit, also shown, corresponds
to a fraction of events with a neutron of $(0.3 \pm 1.7)$\% and thus agrees well with 
expectations. 

The sample of inclusive electron events discussed in 
Section  VI was also analyzed for neutron production.
To enhance the sensitivity to possible sources with neutrons, the requirement
$\rm{cos}~\theta < 0.9$ was imposed.  This eliminated most neutrino electron elastic
scattering events.   Similarly, events with identified $\beta$ decays were excluded.
Fig.~24(a) shows the R distribution for the beam-excess sample.  The fraction of events
with neutrons obtained from the fit to this distribution is $(-3.4 \pm 2.8)$\%.
Thus there cannot be a significant 
background in the $^{12}\rm{C}(\nu_e, e^-)^{12}\rm{N}^*$ 
sample due to any source of events with associated neutrons.

A similar analysis for events in the beam-off data sample provides an
improved understanding of the sources of cosmic ray backgrounds.  High energy
cosmic ray neutrons which enter the detector will occasionally produce ``electron like''
events which satisfy the electron particle ID criteria.  The magnitude of this
neutron background is very sensitive to the requirement on $\chi_{tot}$ as
can be seen in Fig.~3.  For the physics analysis in this paper the relatively
loose criteria $\chi_{tot} < 0.85$ has been used.  The cosmic ray neutron component
can be independently determined from the R distribution since the neutrons 
eventually thermalize and produce capture $\gamma$s. 
Fig.~24(b) shows the R distribution for the beam-off sample. The
fraction of events with neutrons obtained from the fit to this distribution is
$(11.8 \pm 0.6)$\%. If we apply a tighter particle ID criteria, $\chi_{tot} < 0.65$, the
neutron component is reduced to $(1.6 \pm 0.5)$\%.

Thus the cosmic ray neutron component can be identified both from the particle ID 
parameter, $\chi_{tot}$, obtained from the fit to the primary event and from the $\gamma$
likelihood ratio R.
Fig.~25(a)  
shows the $\chi_{tot}$ distribution for all beam-off events in the sample.
The $\chi_{tot}$ distribution for events with R$>$ 30, shown in Fig.~25(b)  , is very
different.  These events should arise predominately from cosmic ray neutron interactions
and, indeed, their $\chi_{tot}$ distribution is very similar to the distribution for
neutrons shown in Fig.~3 for the region of concern, $\chi_{tot}<0.85$.
A better procedure than simply requiring R$>$ 30 is to use the R information to
extract the $\chi_{tot}$ distributions for the correlated and uncorrelated components.
The $\chi_{tot}$ distribution for the uncorrelated component, shown in Fig.~25(c), is similar to
that obtained from Michel electrons except for a small excess 
at high $\chi_{tot}$.
The $\chi_{tot}$ distribution for 
the correlated component, shown in Fig.~25(d), agrees with the $\chi_{tot}$ distribution
expected for cosmic ray neutrons.
Thus there are two main types of cosmic ray events that pass the selection criteria:
(1) events due to electron or photon interactions followed only by uncorrelated photons and 
(2) events due to cosmic ray neutron interactions.  Further, the cosmic ray neutron component
can be reduced to a low level by applying a tight particle ID requirement as is done in
the oscillation search~\cite{Bigpaper2}.
This study of the beam-off events is useful as a test of the analysis technique and 
our understanding of the backgrounds and detector response.
It should be emphasized, however, that the physics analysis presented in 
sections V and VI depends  on the highly reliable beam-off subtraction procedure to remove
the cosmic ray background that survives the event selection criteria.

\section{Conclusions}

The process $^{12}\rm{C}(\nu_e,e^-)^{12}\rm{N}_{g.s.}$ has been measured with a clean 
sample of 500 events for which both the $e^-$ and the $e^+$ from the beta decay of the 
$^{12}\rm{N}_{g.s.}$ are detected. 
For this process cross section calculations using empirical form factors 
are expected to be very reliable.
The flux averaged cross section is measured to be 
$(9.1 \pm 0.4 \pm 0.9)\times 10^{-42} \rm{cm}^2$ in good agreement with 
other experiments and theoretical expectations.  The angular and
energy distributions of the electron also agree well with theoretical
expectations.

The process $^{12}\rm{C}(\nu_e,e^-)^{12}\rm{N}^*$ has also been measured.
There are larger uncertainties in this calculated inclusive  cross section 
than for the  $^{12}\rm{N}_{g.s.}$ transition.
The flux averaged cross section is found to be 
$(5.7 \pm 0.6 \pm 0.6)\times 10^{-42} \rm{cm}^2$, in agreement with
a recent CRPA calculation and earlier but less precise experimental results as
shown in Table VII.
The energy and angular distributions are 
also consistent with theoretical expectations.

\paragraph*{Acknowledgments}

The authors gratefully acknowledge the support of Peter Barnes,
Cyrus Hoffman, and John McClelland. 
It is particularly pleasing that a number of undergraduate students
from participating institutions were able to contribute to the experiment.
We acknowledge many interesting and helpful discussions with  Edwin Kolbe
and Petr Vogel.
This work was conducted under the auspices of the US Department of Energy,
supported in part by funds provided by the University of California for
the conduct of discretionary research by Los Alamos National Laboratory.
This work is also supported by the National Science Foundation.
We are particularly grateful for the extra effort that was made by these
organizations to provide funds for running the accelerator at the end of
the data taking period in 1995.

\clearpage
\begin{table}
\caption { The electron selection criteria and corresponding
efficiencies for 1994 and 1995 for the reaction $^{12}\rm{C}(\nu_e,e^-)^{12}\rm{N}_{g.s.}$.}
\label{I}
\begin{tabular}{llll} 
Quantity      & Criteria &1994 Eff. &1995 Eff.  \\ 
\tableline
Fiducial Volume & D$>$35.0~cm & $0.850 \pm 0.050$ & $0.850 \pm 0.050$ \\ 
Electron Energy & $16.0< \rm{E_e} < 40.0$~MeV & $0.815 \pm $0.005 & $0.815 \pm $ 0.005\\ 
Particle ID  & $\chi_{tot} \leq0.85$ & $0.907 \pm 0.005$ & $0.887 \pm 0.005$ \\ 
Intime Veto  & $<4$ PMTs & $0.995 \pm 0.005$ & $0.989 \pm 0.005$ \\ 
Past Activity  & $\Delta t_p >20,35 \rm{\mu s} $ & $0.673 \pm 0.005$  & $0.714 \pm 0.005$ \\ 
DAQ Dead time  & &  $0.970 \pm 0.010$ & $0.970 \pm 0.010$ \\ 
\tableline
Total         & &  $0.408 \pm 0.025$ & $0.421 \pm 0.026$ \\ 
\end{tabular}
\end{table}

\begin{table}
\caption { Beta decay $e^+$  selection criteria and corresponding
efficiencies for 1994 and 1995 for the reaction $^{12}\rm{C}(\nu_e,e^-)^{12}\rm{N}_{g.s.}$.}
\label{II}
\begin{tabular}{llll} 
Quantity      & Criteria &1994 Eff. &1995 Eff.  \\ 
\tableline
$\beta$ decay time &$52 \rm{\mu s <t < 45 ms}$ & $0.938 \pm 0.002$ & $0.938 \pm 0.002$ \\ 
Spatial Correlation & $\Delta$r$~<1~m$ & $0.964 \pm 0.020$ & $0.964 \pm 0.020$ \\
PMT Threshold & $>$100  for 1994, $>$75 for 1995  & $0.765 \pm 0.015$ & $0.881 \pm 0.010$ \\  
Fiducial Volume & D$>$0~cm & $0.972 \pm 0.010$ & $0.972 \pm 0.010$ \\ 
Trigger Veto  & $>15.1 \rm{\mu s}$  & $0.815 \pm 0.005$ & $0.815 \pm 0.005$ \\ 
Intime Veto  & $<4$ PMTs & $0.995 \pm 0.001$ & $0.989 \pm 0.001$ \\ 
DAQ Dead time  & &  $0.970 \pm 0.010$ & $0.970 \pm 0.010$ \\ 
\tableline
Total         & &  $0.529 \pm 0.017$ & $0.606 \pm 0.017$ \\ 
\end{tabular}
\end{table}

\begin{table}
\caption { Events, accidental backgrounds, efficiencies, neutrino flux and flux
averaged cross sections with statistcal errors only 
for 1994 and 1995 for the reaction $^{12}\rm{C}(\nu_e,e^-)^{12}\rm{N}_{g.s.}$.}
\label{III}
\begin{tabular}{lcc} 
   &1994  &1995  \\
\tableline 
Beam-on Events     & 241 events & 308 events \\ 
Beam-off Events $\times$ Duty Ratio     & $189 \times 0.08$ events & $312 \times 0.06$ events \\ 
Beam-excess Events     & 226 events & 289 events  \\ 
Accidental Background    & 3.3 events & 4.2 events  \\ 
Efficiency    & 0.216 & 0.256  \\ 
$\nu_e$ flux    & $3.04 \times 10^{13}/\rm{cm}^2$ & $3.43 \times 10^{13}/\rm{cm}^2$ \\
$<\sigma>$    & $(9.3 \pm 0.7) \times 10^{-42}\rm{cm}^2$ & $(8.9 \pm 0.6) \times 10^{-42}\rm{cm}^2$ \\  
\end{tabular}
\end{table}

\begin{table}
\caption { Measurements and theoretical predictions of 
the flux averaged cross section for the process 
 $^{12}\rm{C}(\nu_e,e^-)^{12}\rm{N}_{g.s.}$.}
\label{IV}
\begin{tabular}{ll} 
Experiment  &  \\
\tableline 
LSND    & $(9.1 \pm 0.4 \pm 0.9)\times 10^{-42} \rm{cm}^2$     \\ 
E225[2]    & $(10.5 \pm 1.0 \pm 1.0)\times 10^{-42} \rm{cm}^2$ \\ 
KARMEN[3]  & $(9.1 \pm 0.5 \pm 0.8)\times 10^{-42} \rm{cm}^2$ \\ 
\tableline 
\tableline 
Theory  &  \\
\tableline 
Donnelly[6] & $ 9.4 \times 10^{-42} \rm{cm}^2$ \\ 
Fukugita et al.[5] & $ 9.2 \times 10^{-42} \rm{cm}^2$ \\ 
Kolbe et al.[9] & $ 9.3 \times 10^{-42} \rm{cm}^2$ \\ 
Mintz et al.[7] & $ 8.0 \times 10^{-42} \rm{cm}^2$ \\ 
\end{tabular}
\end{table}

\begin{table}
\caption { The electron selection criteria and corresponding
efficiencies for 1994 and 1995 for the reaction $^{12}\rm{C}(\nu_e,e^-)^{12}\rm{N}^*$.}
\label{V}
\begin{tabular}{llll} 
Quantity      & Criteria &1994 Eff. &1995 Eff.  \\ 
\tableline
Fiducial Volume & D$>$35.0~cm & $0.850 \pm 0.050$ & $0.850 \pm 0.050$ \\ 
Vertical Position & $y>$ -120.0~cm & $0.915 \pm 0.015$ & $0.915 \pm 0.015$ \\
Direction Angle & cos~$\theta < 0.9$ & $0.985 \pm 0.010$ & $0.985 \pm 0.010$ \\
Electron Energy & $20.0< \rm{E_e} < 34.0$~MeV & $0.582 \pm $ 0.006& $0.582 \pm $ 0.006\\ 
Particle ID  & $\chi_{tot} \leq0.85$ & $0.908 \pm 0.005$ & $0.888 \pm 0.005$ \\ 
Intime Veto  & $<4$ PMTs & $0.995 \pm 0.005$ & $0.989 \pm 0.005$ \\ 
Past Activity  & $\Delta t_p >20,35 \rm{\mu s} $ & $0.673 \pm 0.005$ & $0.714 \pm 0.005$ \\
DAQ Dead time  & &  $0.970 \pm 0.010$ & $0.970 \pm 0.010$ \\ 
\tableline
Total         & &  $0.263 \pm 0.017 $ & $0.271 \pm 0.018$ \\ 
\end{tabular}
\end{table}

\begin{table}
\caption { Number of observed events, calculated background events and events
 attributed to the reaction $^{12}\rm{C}(\nu_e,e^-)^{12}\rm{N}^*$. The flux 
averaged cross sections with statistical errors are also shown.}
\label{VI}
\begin{tabular}{lcc} 
   &1994  &1995  \\ 
\tableline
Beam-on Events     & 695 events & 689 events  \\ 
Beam-excess Events     & 302.5 events& 357.8 events \\ 
$\nu e$ elastic background  & 4.2 events & 4.8 events  \\ 
$^{13}$C background  & $12.2 \pm 6.1$ events & $14.3 \pm 7.1$ events  \\ 
$^{12}\rm{N}_{g.s.}$ background & $141.6 \pm 11.4$ events & $122.0 \pm 9.0$ events \\
$^{12}\rm{C}(\nu_e,e^-)^{12}\rm{N}^*$ & $144.5 \pm 28.8 $ events & $216.7 \pm 27.7$ events \\
$<\sigma>$    & $(5.0 \pm 1.0) \times 10^{-42}\rm{cm}^2$ & $(6.4 \pm 0.8) \times 10^{-42}\rm{cm}^2$ \\  
\end{tabular}
\end{table}

\begin{table}
\caption { Measurements and theoretical predictions of 
the flux averaged cross section for the process 
$^{12}\rm{C}(\nu_e,e^-)^{12}\rm{N}^*$.}
\label{VII}
\begin{tabular}{ll} 
Experiment  &  \\
\tableline 
LSND    & $(5.7 \pm 0.6 \pm 0.6)\times 10^{-42} \rm{cm}^2$     \\ 
E225[2]    & $ (5.4 \pm 1.9)\times 10^{-42} \rm{cm}^2$ \\ 
KARMEN[3]  & $(6.1 \pm 0.9 \pm 1.0)\times 10^{-42} \rm{cm}^2$ \\ 
\tableline 
\tableline 
Theory  &  \\
\tableline 
Donnelly[6] & $ 3.7 \times 10^{-42} \rm{cm}^2$ \\ 
Kolbe et al.[9] & $ 6.3 \times 10^{-42} \rm{cm}^2$ \\ 
\end{tabular}
\end{table}

%
%

\clearpage

\begin{figure}
\centerline{\psfig{figure=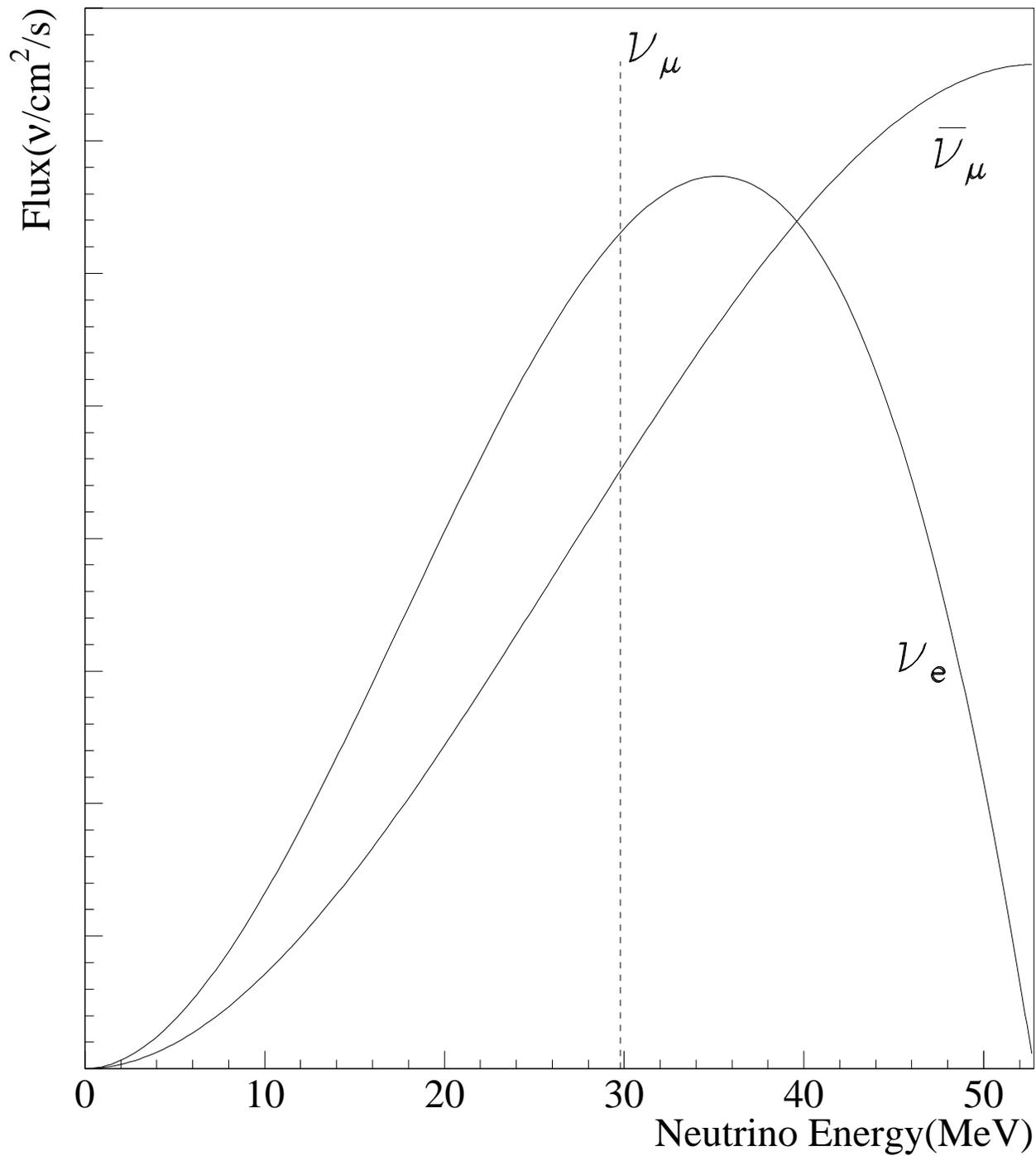,width=7.5in,silent=}}
\caption{Flux shape of neutrinos from pion and muon decay at rest.}
\label{Fig. 1}
\end{figure}

\begin{figure}
\centerline{\psfig{figure=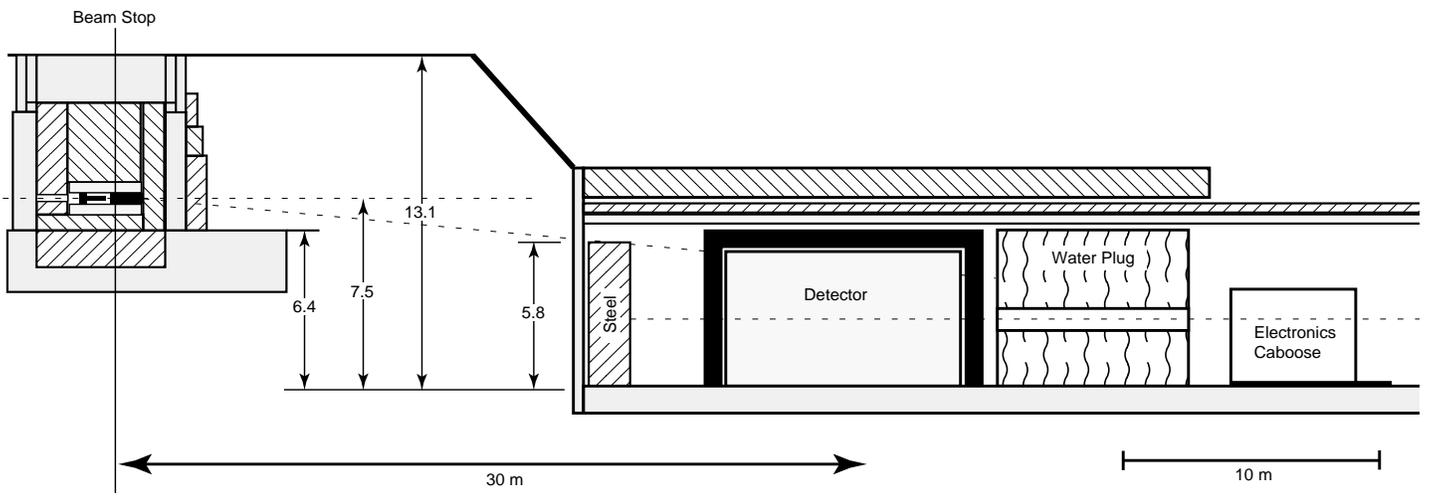,width=7.5in,silent=}}
\caption{Detector enclosure and target area configuration, elevation view.}
\label{Fig. 2}
\end{figure}

\begin{figure}
\centerline{\psfig{figure=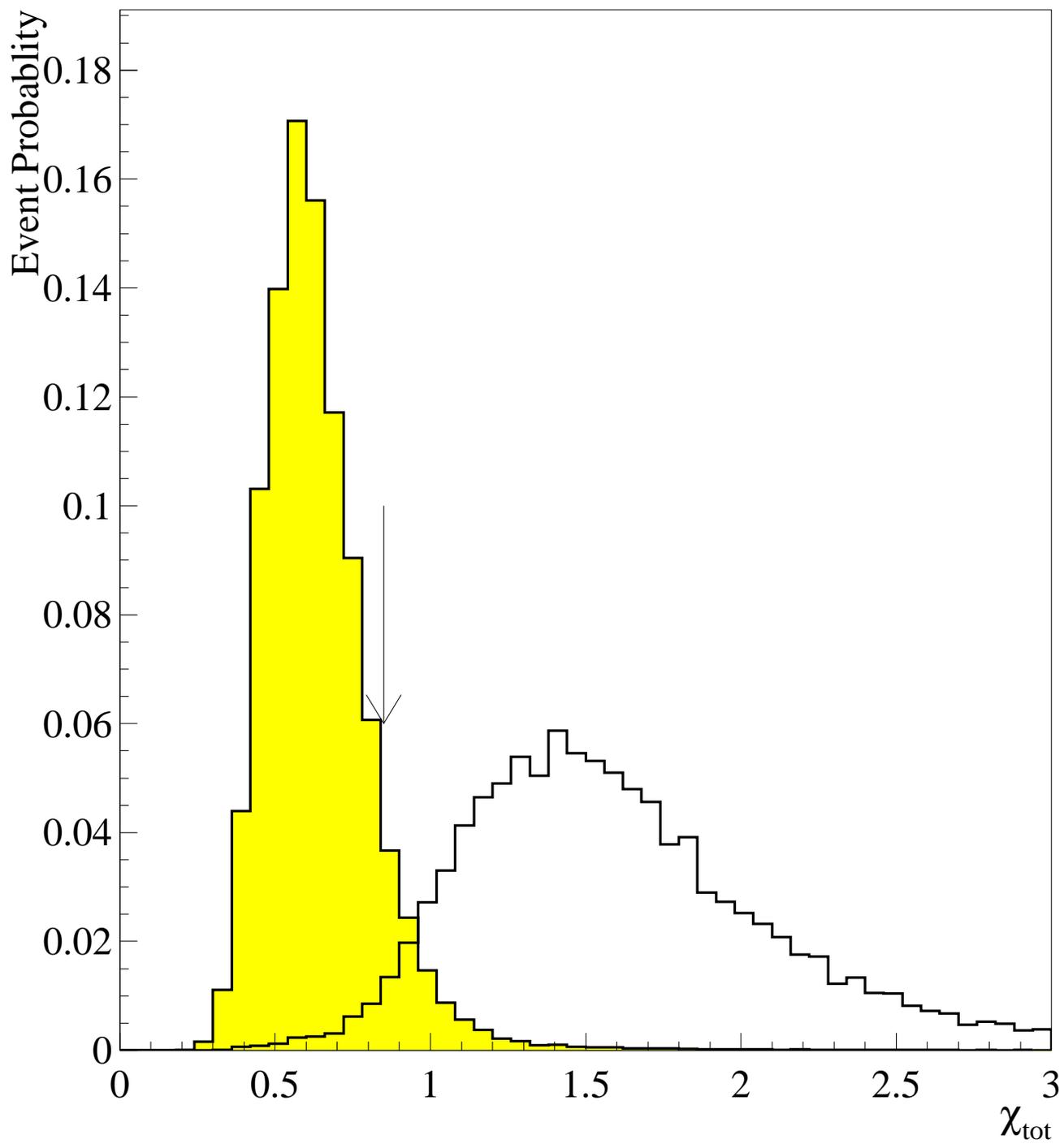,width=7.5in,silent=}}
\caption{Particle ID parameter for ``electrons'' (shaded) and ``neutrons''.}
\label{Fig. 3}
\end{figure}

\begin{figure}
\centerline{\psfig{figure=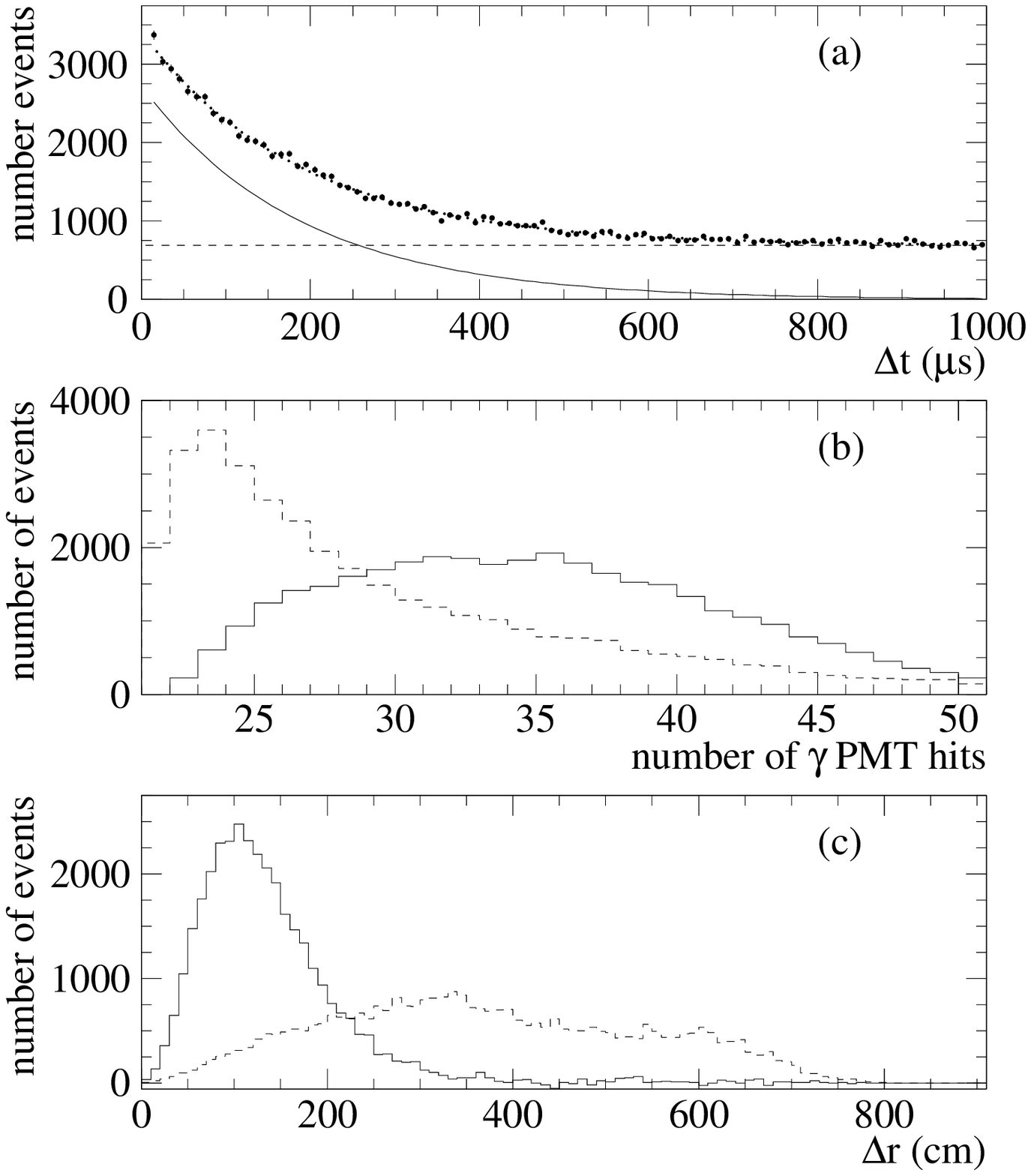,width=7.5in,silent=}}
\caption{Distributions obtained from cosmic ray neutron data for
$\gamma$s that are correlated (solid) or uncorrelated (dashed)
with the primary event: (a) the time
between the photon and primary event; (b) the number of photon PMT hits;
and (c) the distance between the photon and primary event. The raw data points
are also shown in (a).}
\label{Fig. 4}
\end{figure}

\begin{figure}
\centerline{\psfig{figure=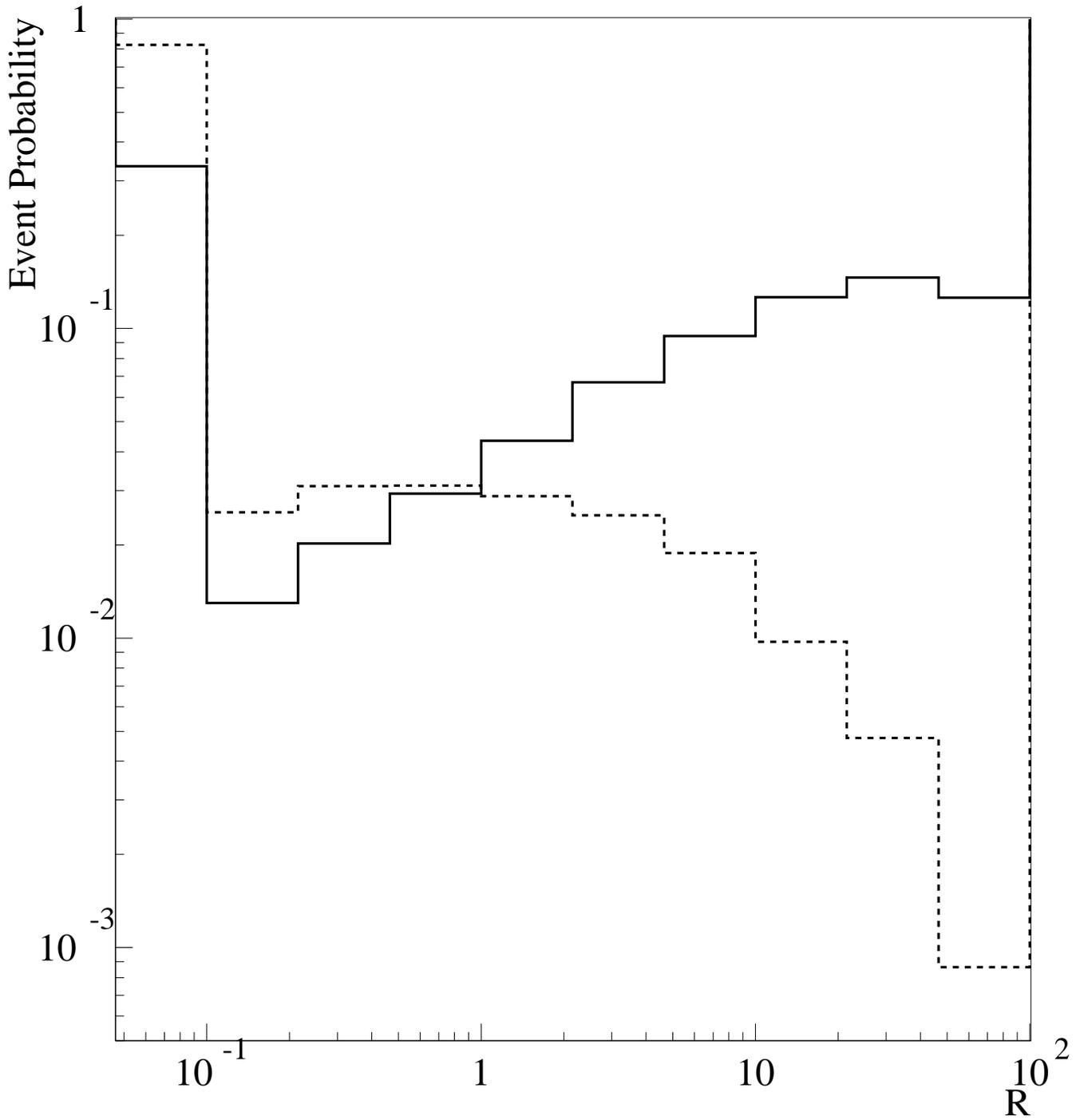,width=7.5in,silent=}}
\caption{Measured R distribution for events
with the $\gamma$ correlated (solid) and uncorrelated (dashed)
with the primary event. 
}
\label{Fig. 5}
\end{figure}

\begin{figure}
\centerline{\psfig{figure=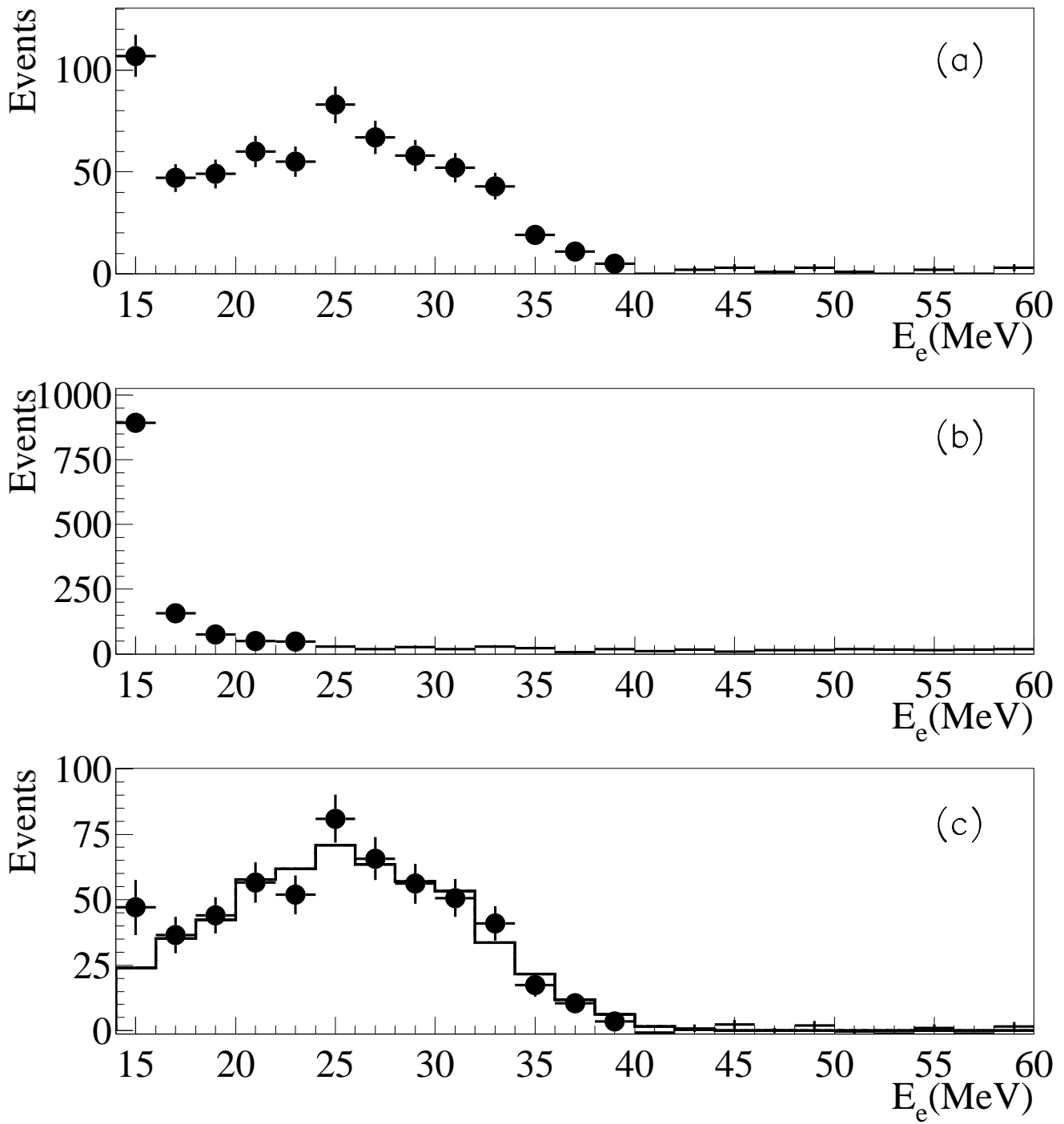,width=7.5in,silent=}}
\caption{Measured electron energy spectrum for (a) beam-on events, (b) beam-off events and
(c) beam-excess events compared with with the expected (solid 
 line) distribution.
  An $e^\pm$ in delayed coincidence is required.}
\label{Fig. 6}
\end{figure}

\begin{figure}
\centerline{\psfig{figure=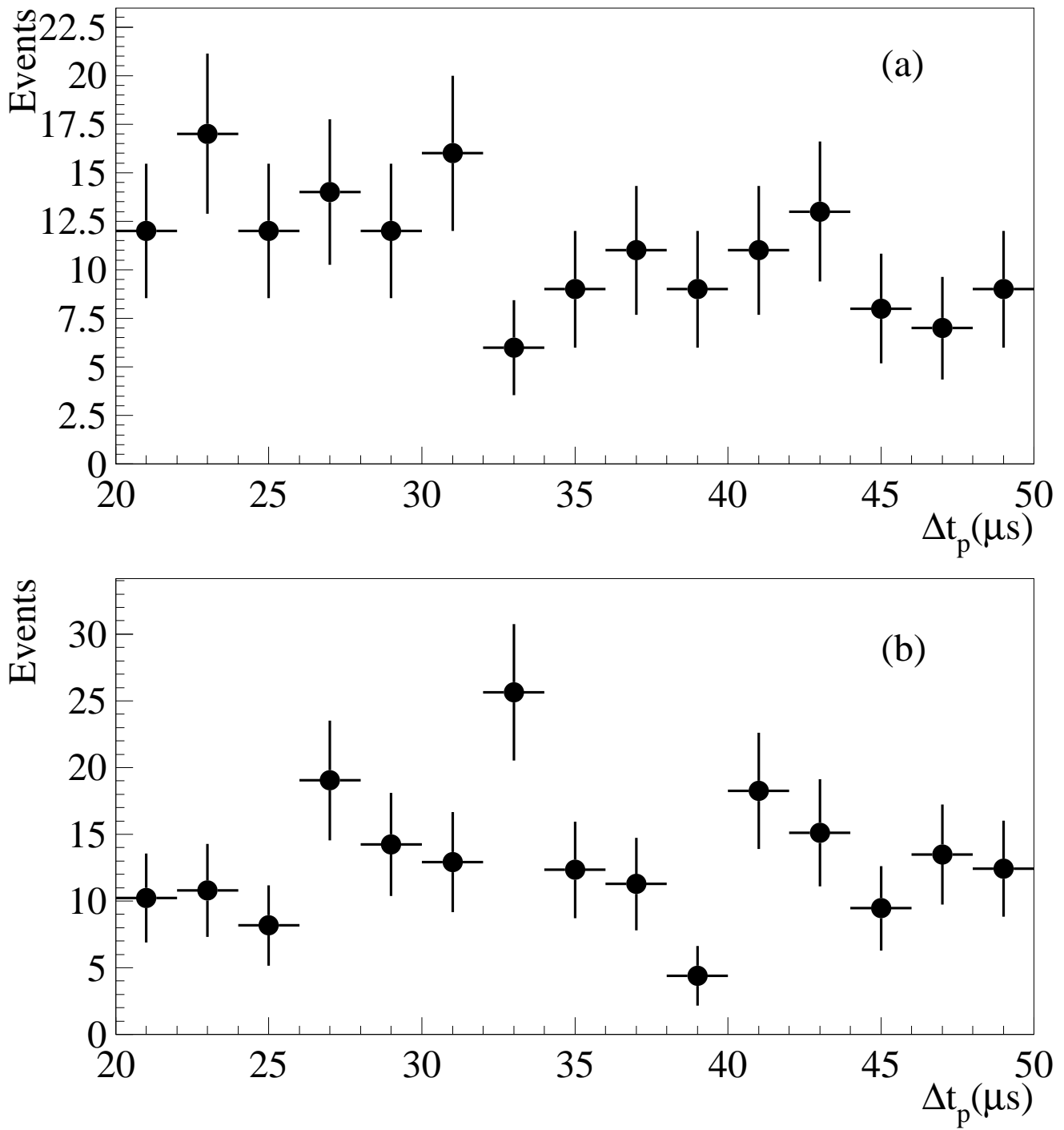,width=7.5in,silent=}}
\caption{Distribution of $\Delta t_p$, the time to past activities,
for (a) beam-off events and (b) beam-excess events.}
\label{Fig. 7}
\end{figure}

\begin{figure}
\centerline{\psfig{figure=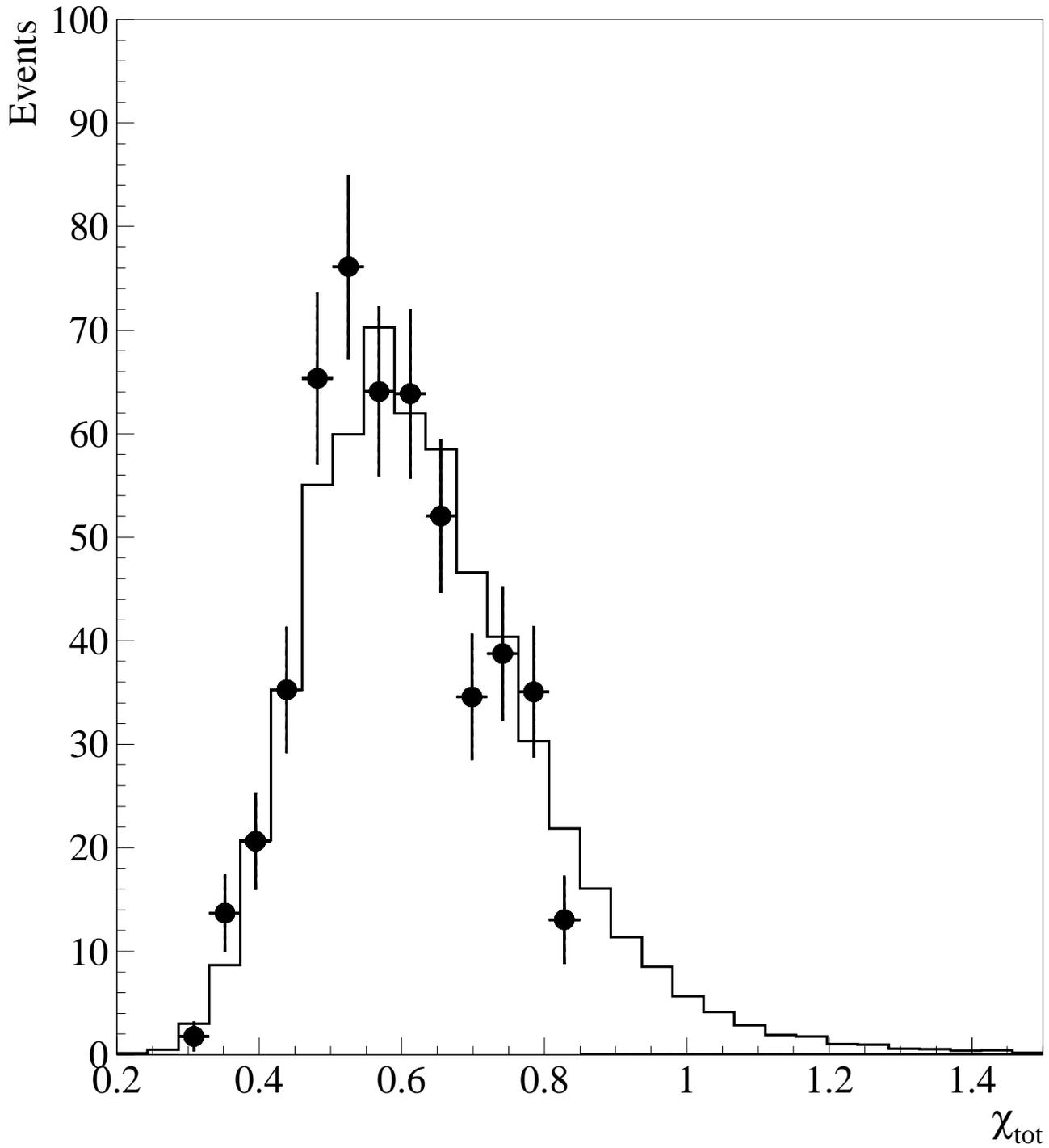,width=7.5in,silent=}}
\caption{Distribution of the particle ID parameter, $\chi_{tot}$, for 
$e^-$ compared with the distribution obtained from Michel electrons (solid line).} 
\label{Fig. 8}
\end{figure}

\begin{figure}
\centerline{\psfig{figure=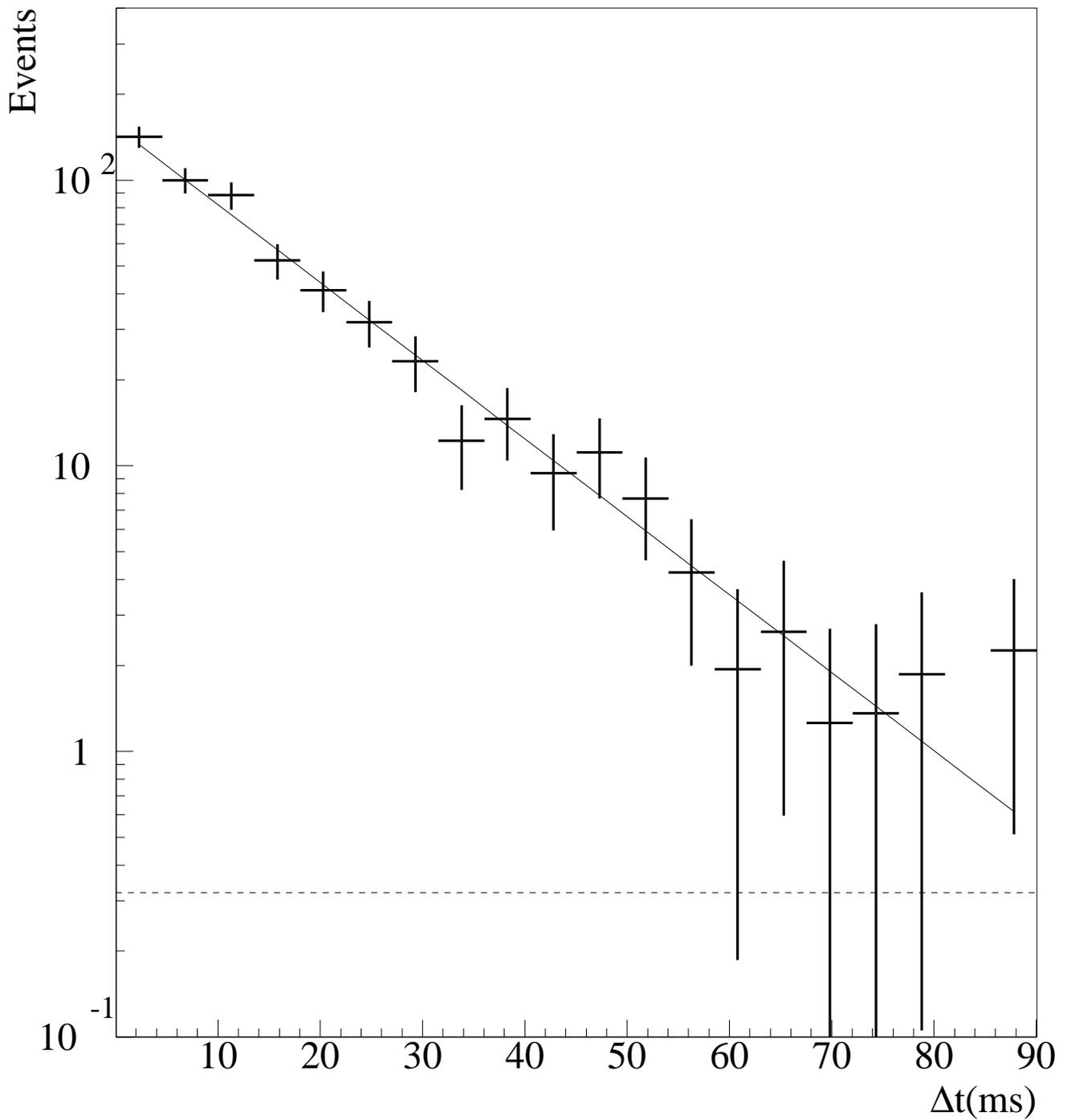,width=7.5in,silent=}}
\caption{Distribution of time between the $e^-$ and $e^+$ for beam-excess 
events in the $^{12}\rm{C}(\nu_e,e^-)^{12}\rm{N}_{g.s.}$ sample. 
The expected distribution is shown with the solid line. The calculated
accidental contribution is shown by the dashed line.}
\label{Fig. 9}
\end{figure}

\begin{figure}
\centerline{\psfig{figure=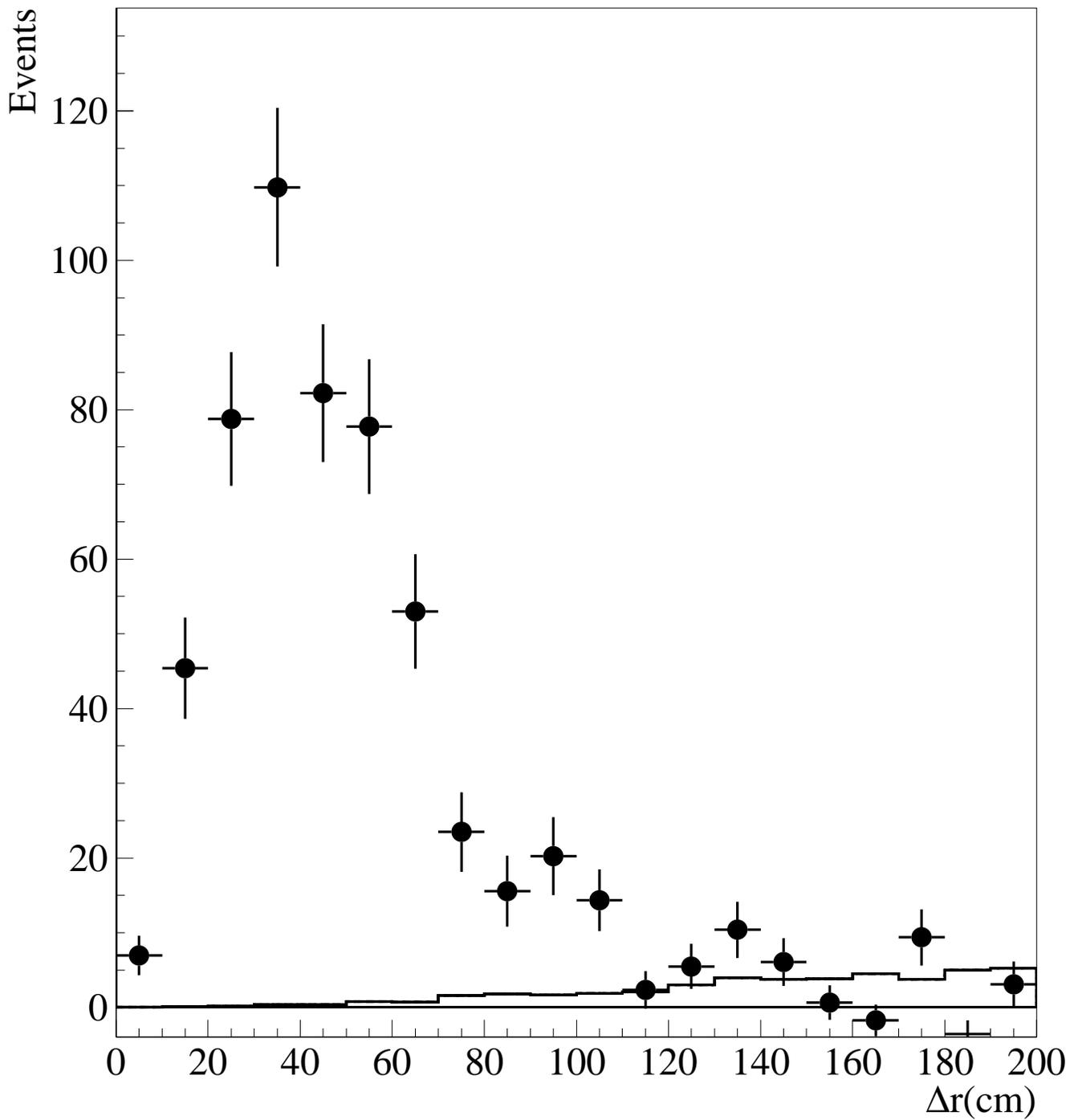,width=7.5in,silent=}}
\caption{Distribution of the distance between reconstructed positions
of $e^-$ and $e^+$ for beam-excess events in the 
$^{12}\rm{C}(\nu_e,e^-)^{12}\rm{N_{g.s.}}$ sample. The calculated accidental
contribution is shown by the solid line.}
\label{Fig. 10}
\end{figure}

\begin{figure}
\centerline{\psfig{figure=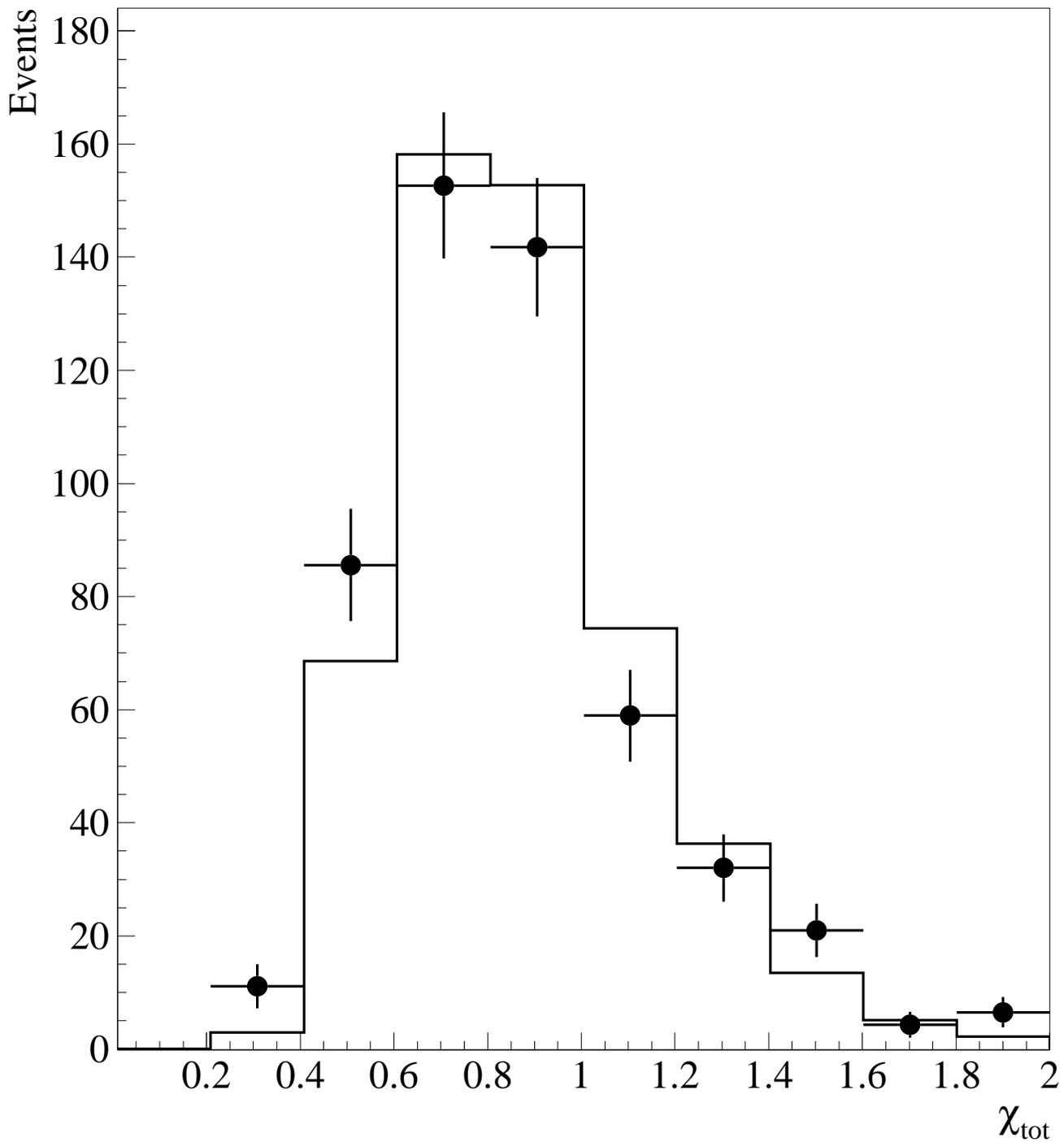,width=7.5in,silent=}}
\caption{Distribution of the particle ID parameter, $\chi_{tot}$, 
for the $e^+$ from the beta decay of $^{12}\rm{N}_{g.s.}$
compared with the expected (solid line) $\chi_{tot}$ distribution.} 
\label{Fig. 11}
\end{figure}

\begin{figure}
\centerline{\psfig{figure=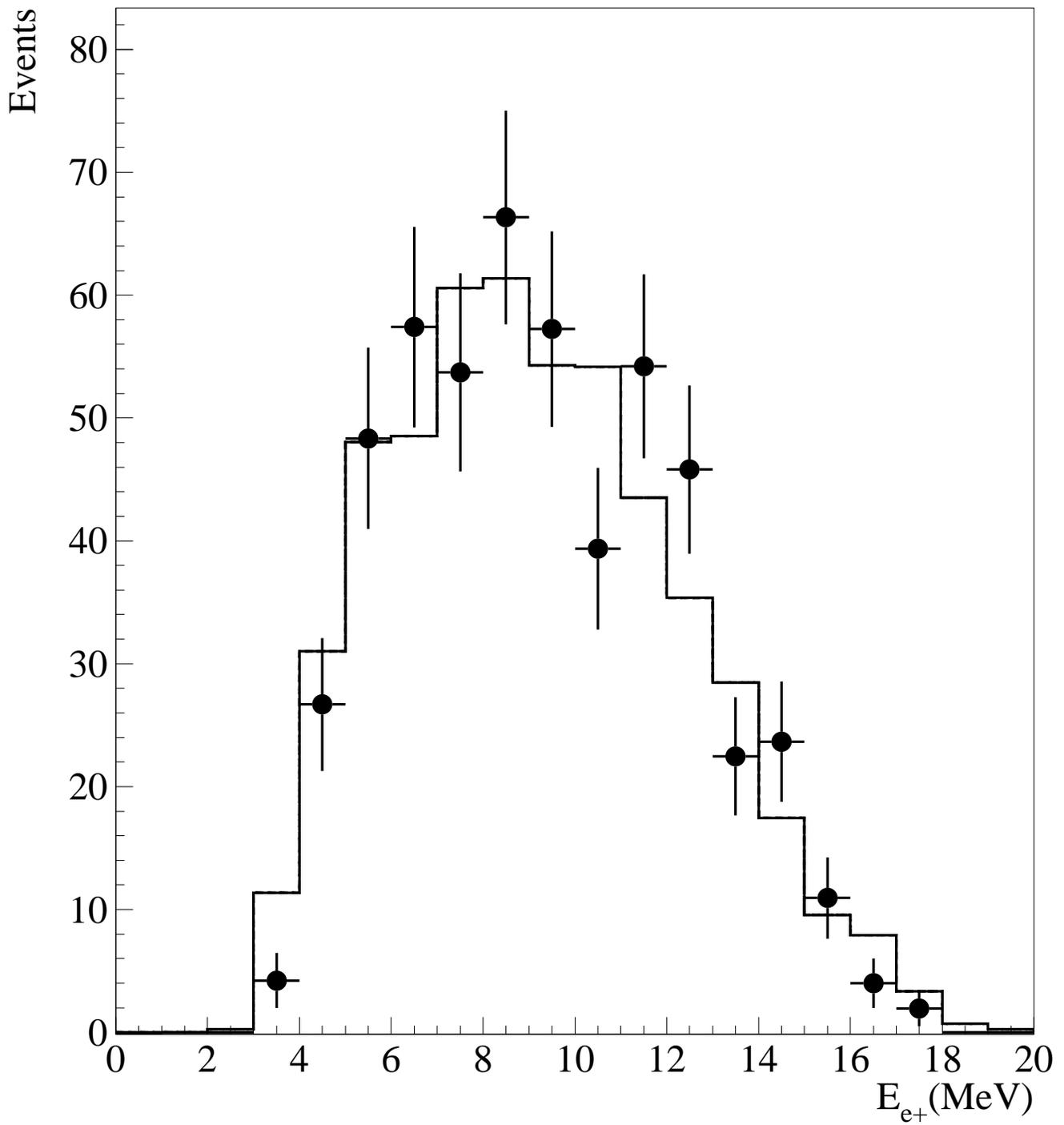,width=7.5in,silent=}}
\caption{Observed and expected (solid line) $e^+$ energy
distribution for events satisfying all selection criteria.}
\label{Fig. 12}
\end{figure}

\begin{figure}
\centerline{\psfig{figure=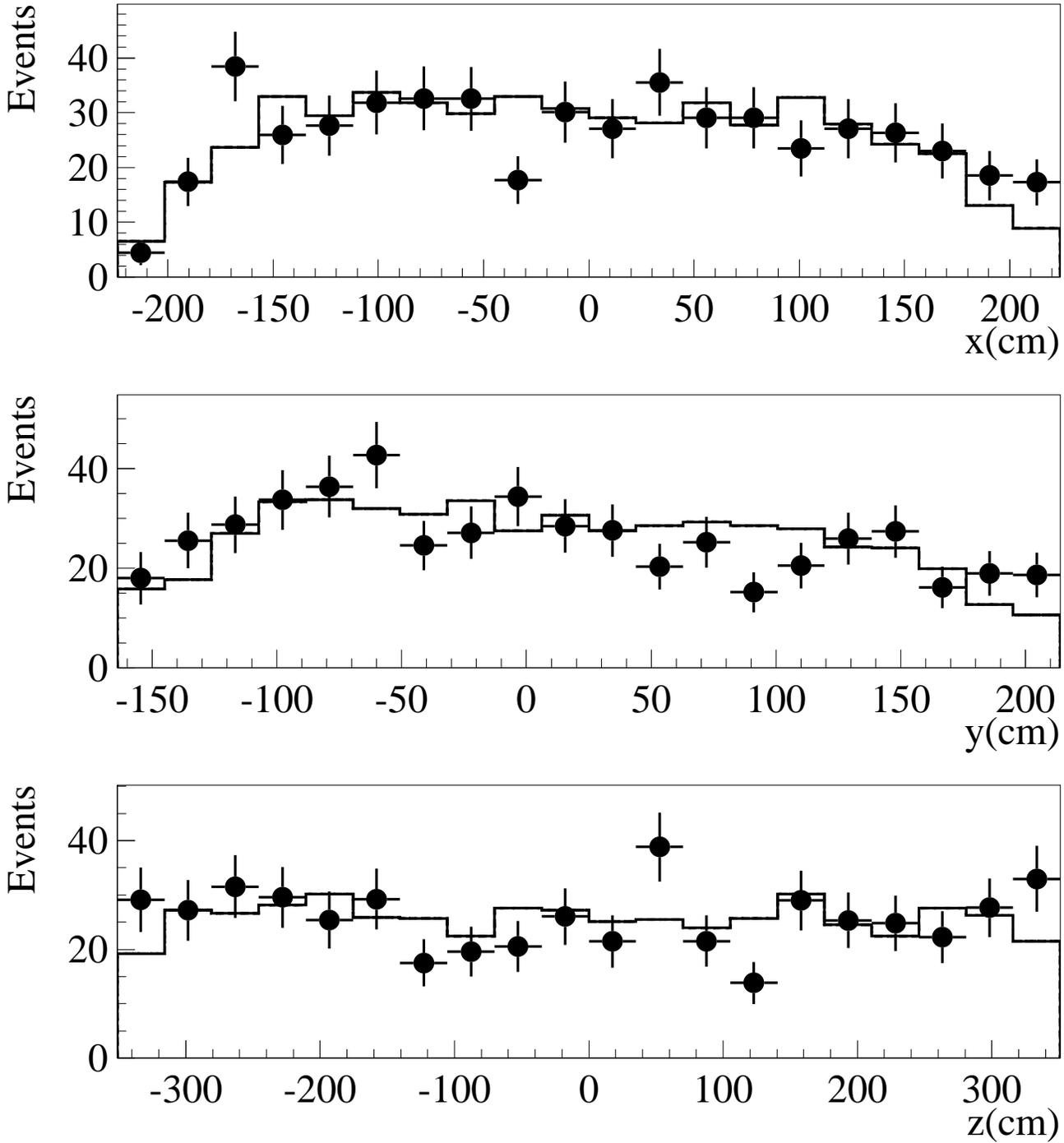,width=7.5in,silent=}}
\caption{The spatial distribution of the $e^-$ for beam-excess 
events in the $^{12}\rm{C}(\nu_e,e^-)^{12}\rm{N}_{g.s.}$ sample compared with 
expectations (solid line) from LSNDMC.}
\label{Fig. 13}
\end{figure}

\begin{figure}
\centerline{\psfig{figure=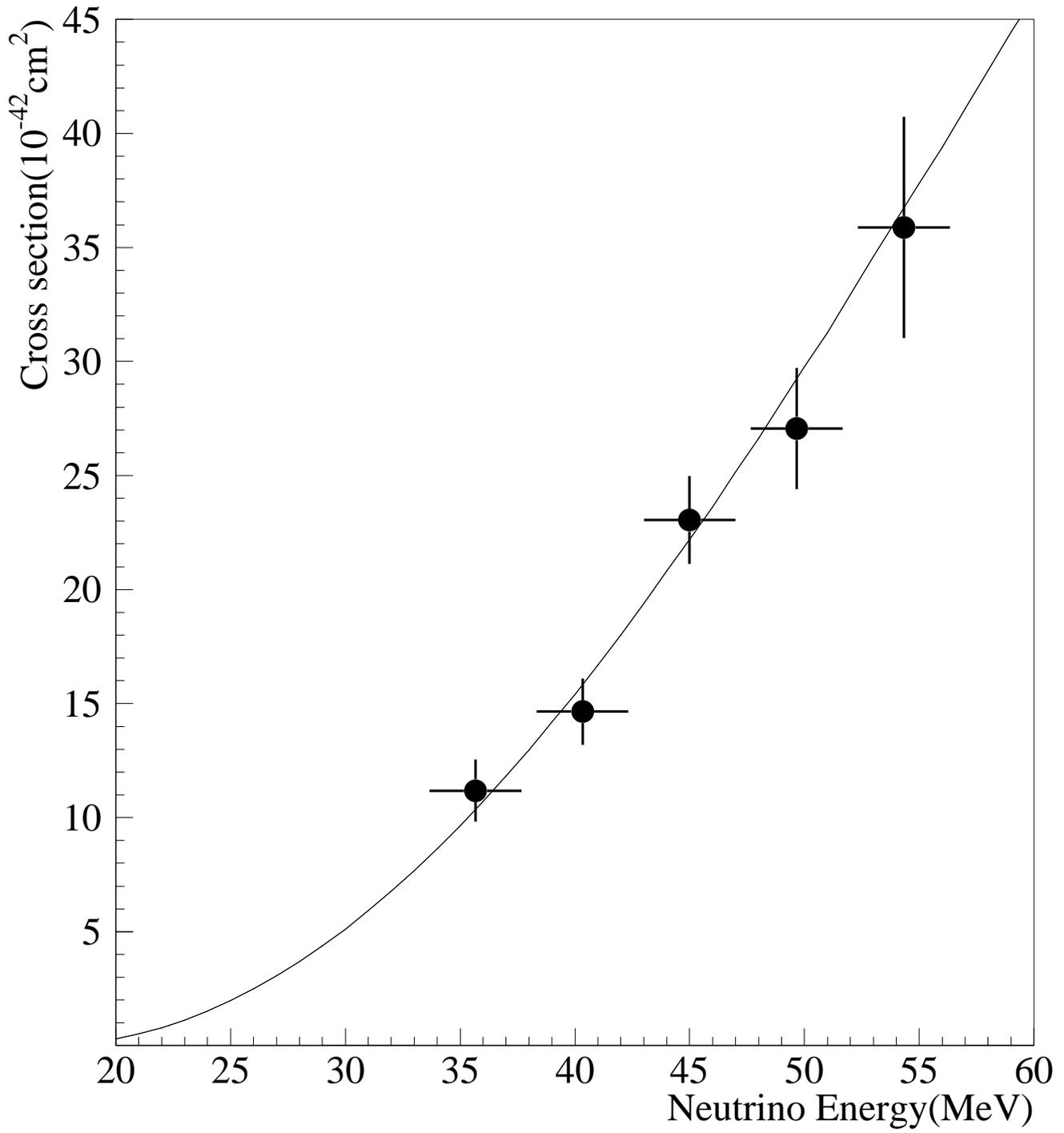,width=7.5in,silent=}}
\caption{The measured and expected (solid line) cross section for the process 
$^{12}\rm{C}(\nu_e,e^-)^{12}\rm{N}_{g.s.}$.
} 
\label{Fig. 14}
\end{figure}

\begin{figure}
\centerline{\psfig{figure=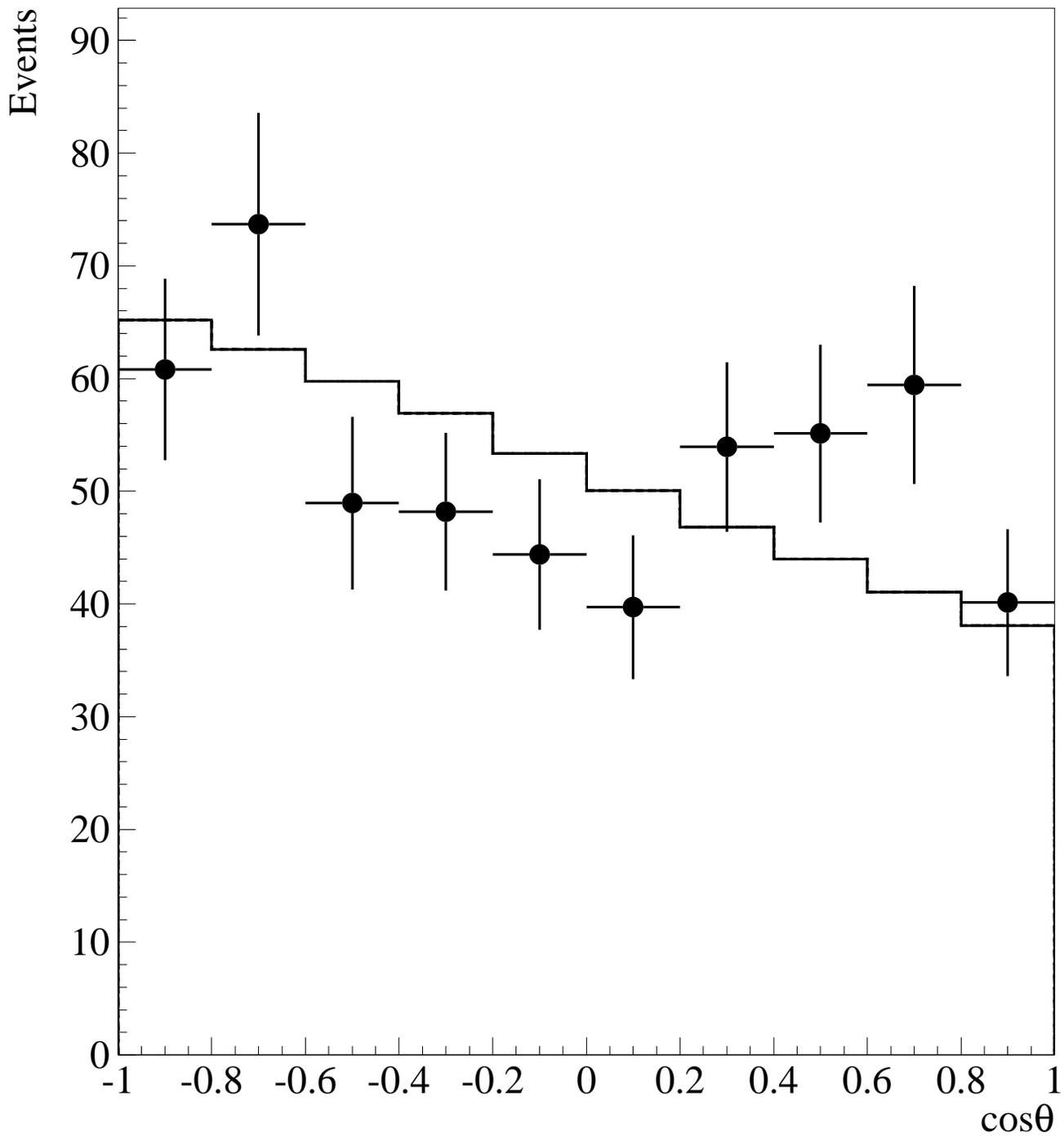,width=7.5in,silent=}}
\caption{Observed and expected (solid line) distribution in 
cos~$\theta$ for the $^{12}\rm{C}(\nu_e,e^-)^{12}\rm{N}_{g.s.}$ sample,
where $\theta$ is the angle between the $e^-$ and the incident
neutrino. }
\label{Fig. 15}
\end{figure}

\begin{figure}
\centerline{\psfig{figure=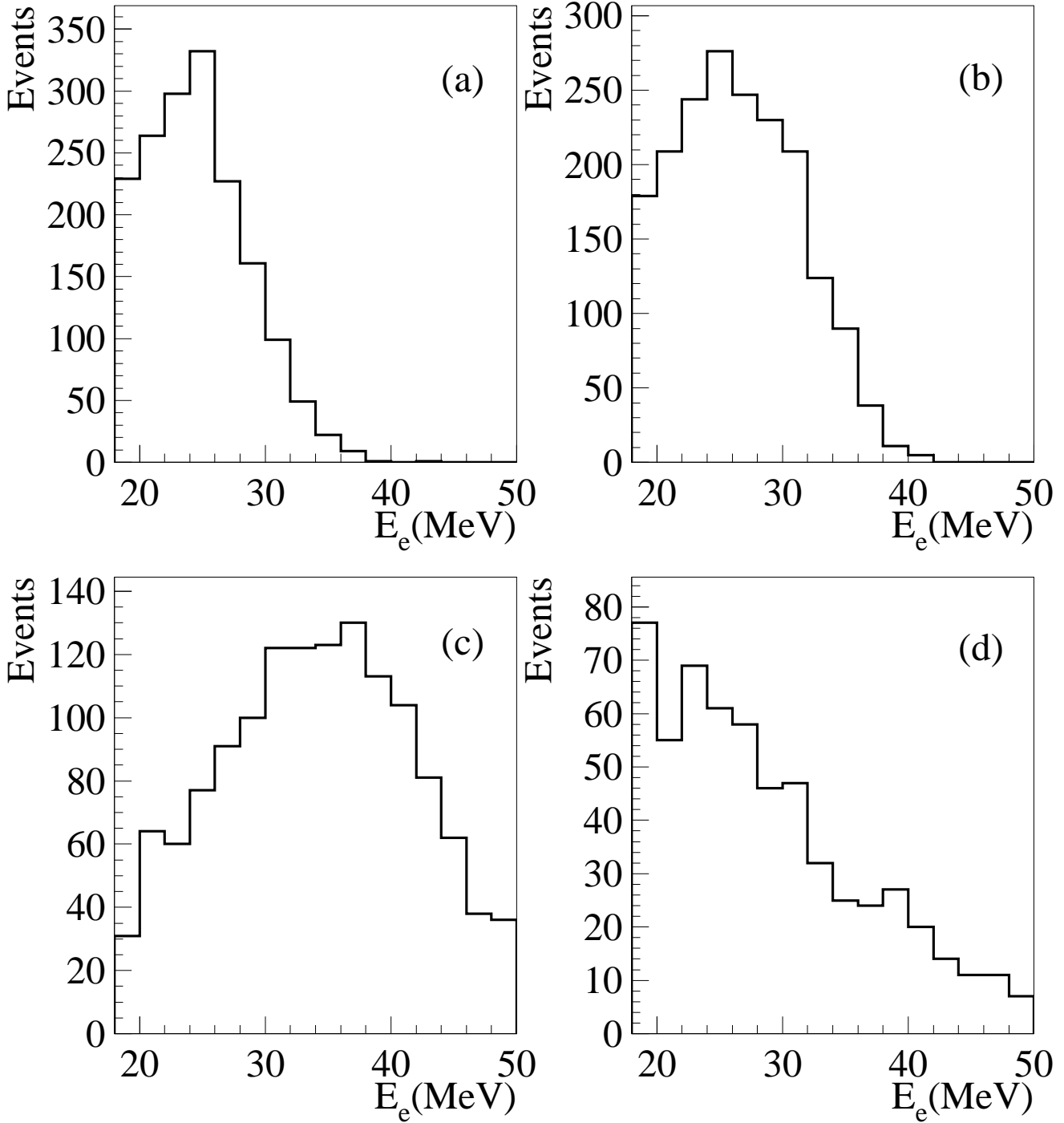,width=7.5in,silent=}}
\caption{Electron energy distribution expected for 
(a) $^{12}\rm{C}(\nu_e, e^-)^{12}\rm{N}^*$, 
(b) $^{12}\rm{C}(\nu_e, e^-)^{12}\rm{N}_{g.s.}$, 
(c) $^{13}\rm{C}(\nu_e, e^-)^{13}\rm{X}$ and 
(d) $\nu$e elastic scattering.
}
\label{Fig. 16}
\end{figure}

\begin{figure}
\centerline{\psfig{figure=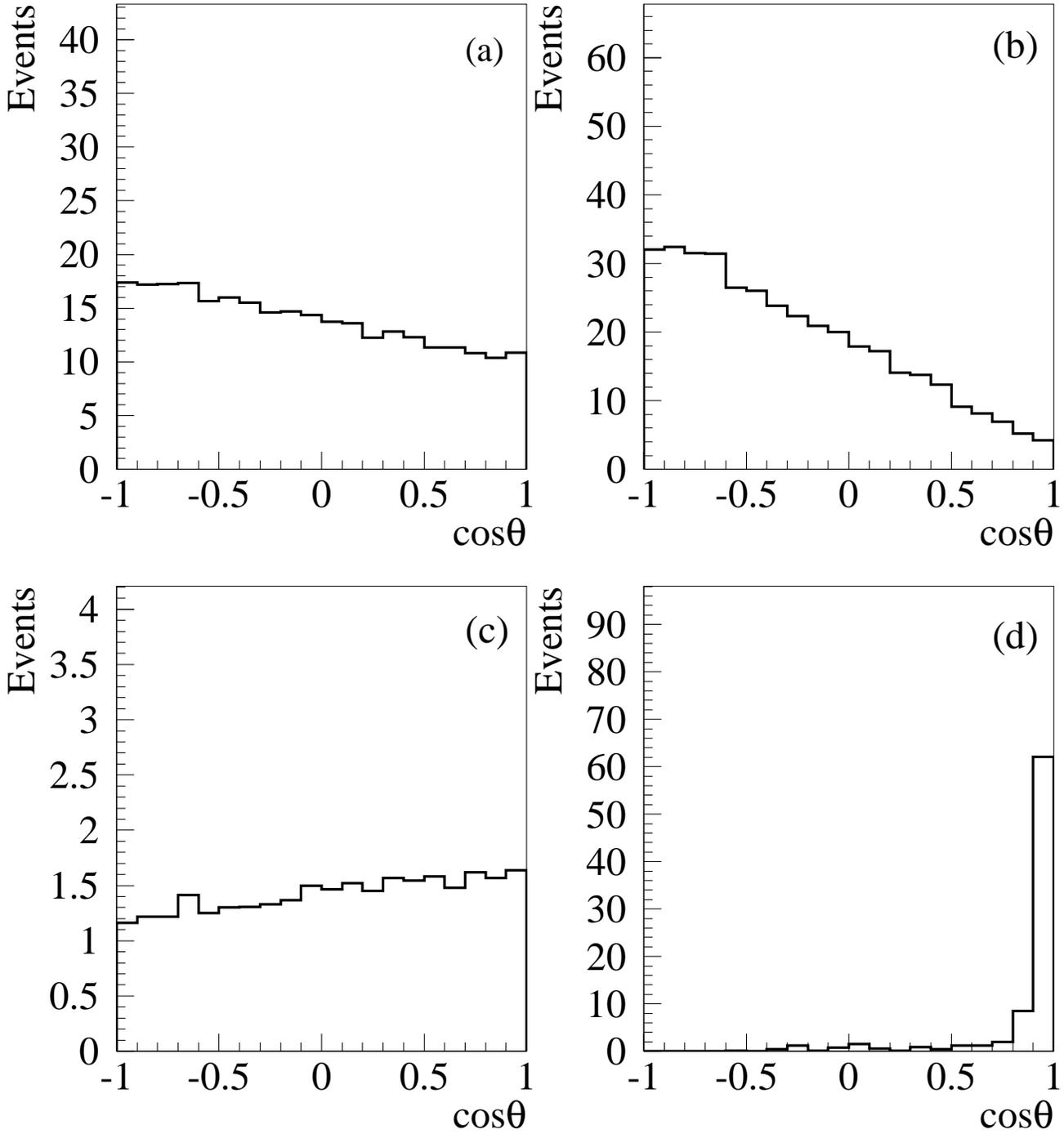,width=7.5in,silent=}}
\caption{Expected distribution in $\rm{cos~\theta}$ for  \newline
(a) $^{12}\rm{C}(\nu_e, e^-)^{12}\rm{N}_{g.s.}$, 
(b) $^{12}\rm{C}(\nu_e, e^-)^{12}\rm{N}^*$, \newline
(c) $^{13}\rm{C}(\nu_e, e^-)^{13}\rm{X}$ and 
(d) $\nu$e elastic scattering.}
\label{Fig. 17}
\end{figure}

\begin{figure}
\centerline{\psfig{figure=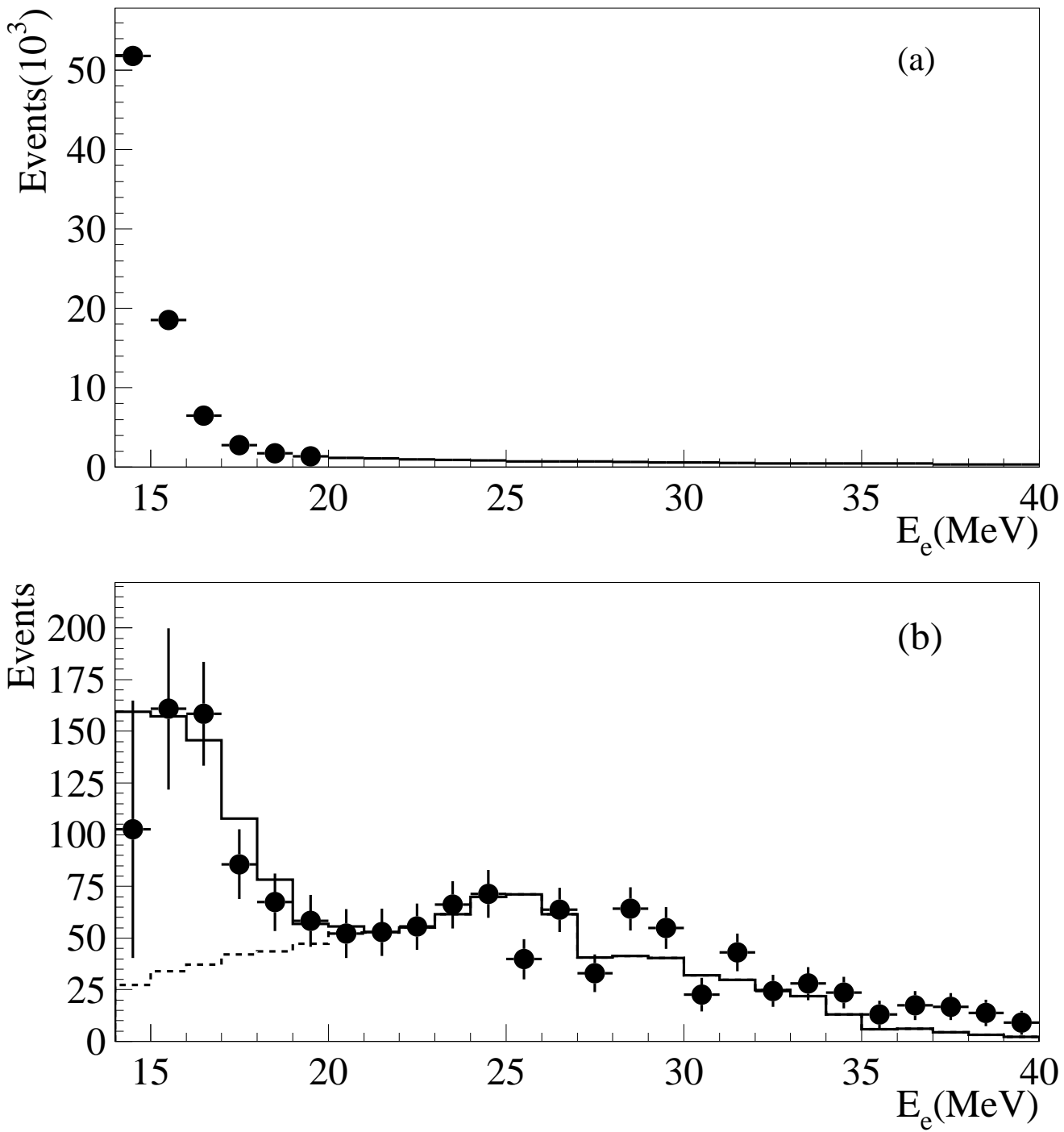,width=7.5in,silent=}}
\caption{Observed electron energy distribution for (a)
beam-off events (b) beam-excess events.
Identified $^{12}\rm{N}_{g.s.}$ events are excluded.
The solid (dashed) line shows the expected distribution including (excluding)
 the 15.1 MeV $\gamma$ contribution from the neutral current excitation of 
$^{12}$C. } 
\label{Fig. 18}
\end{figure}

\begin{figure}
\centerline{\psfig{figure=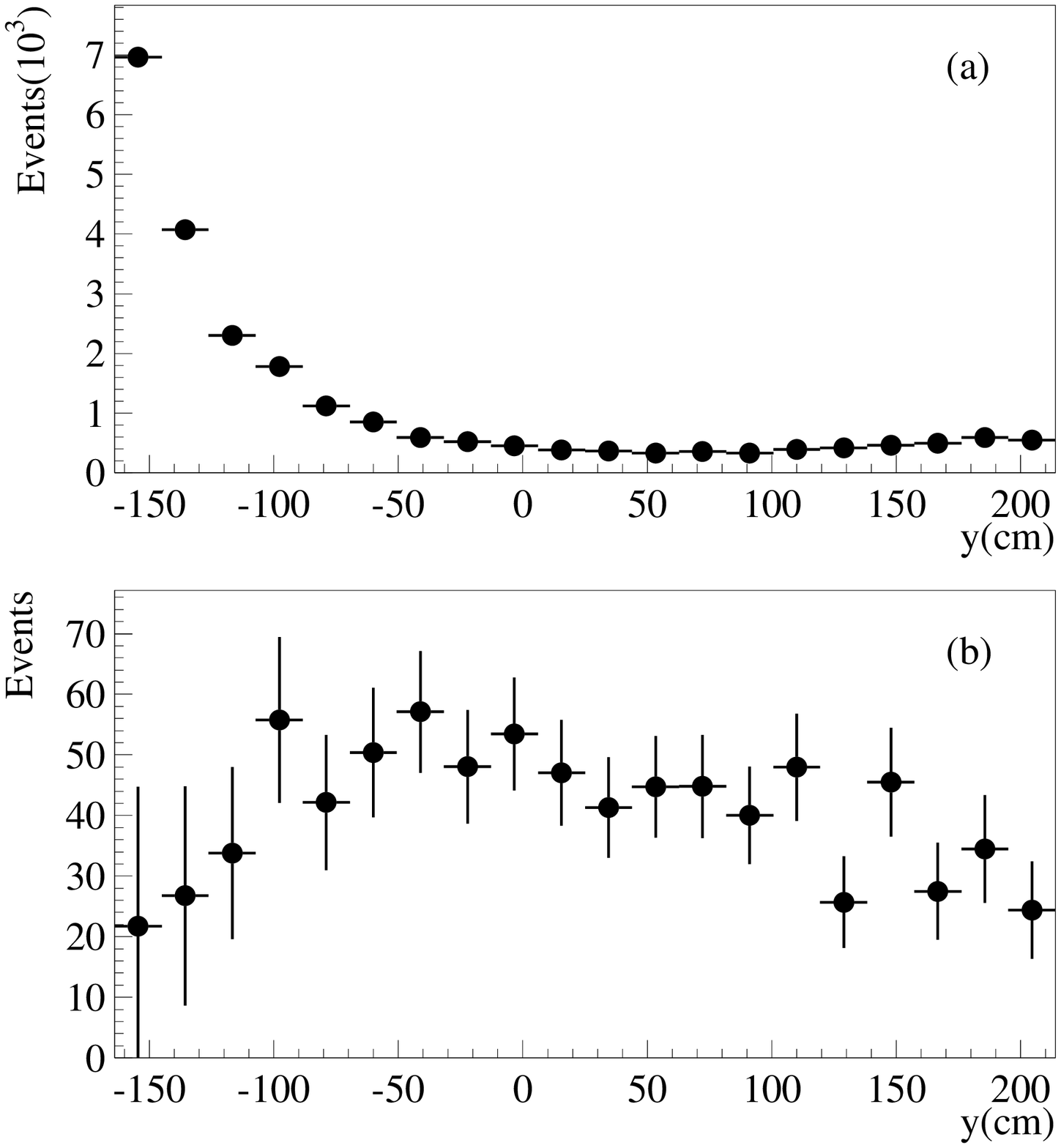,width=7.5in,silent=}}
\caption{Distribution of vertical position $y$ for  (a)
beam-off events and (b) beam-excess events.
}
\label{Fig. 19}
\end{figure}

\begin{figure}
\centerline{\psfig{figure=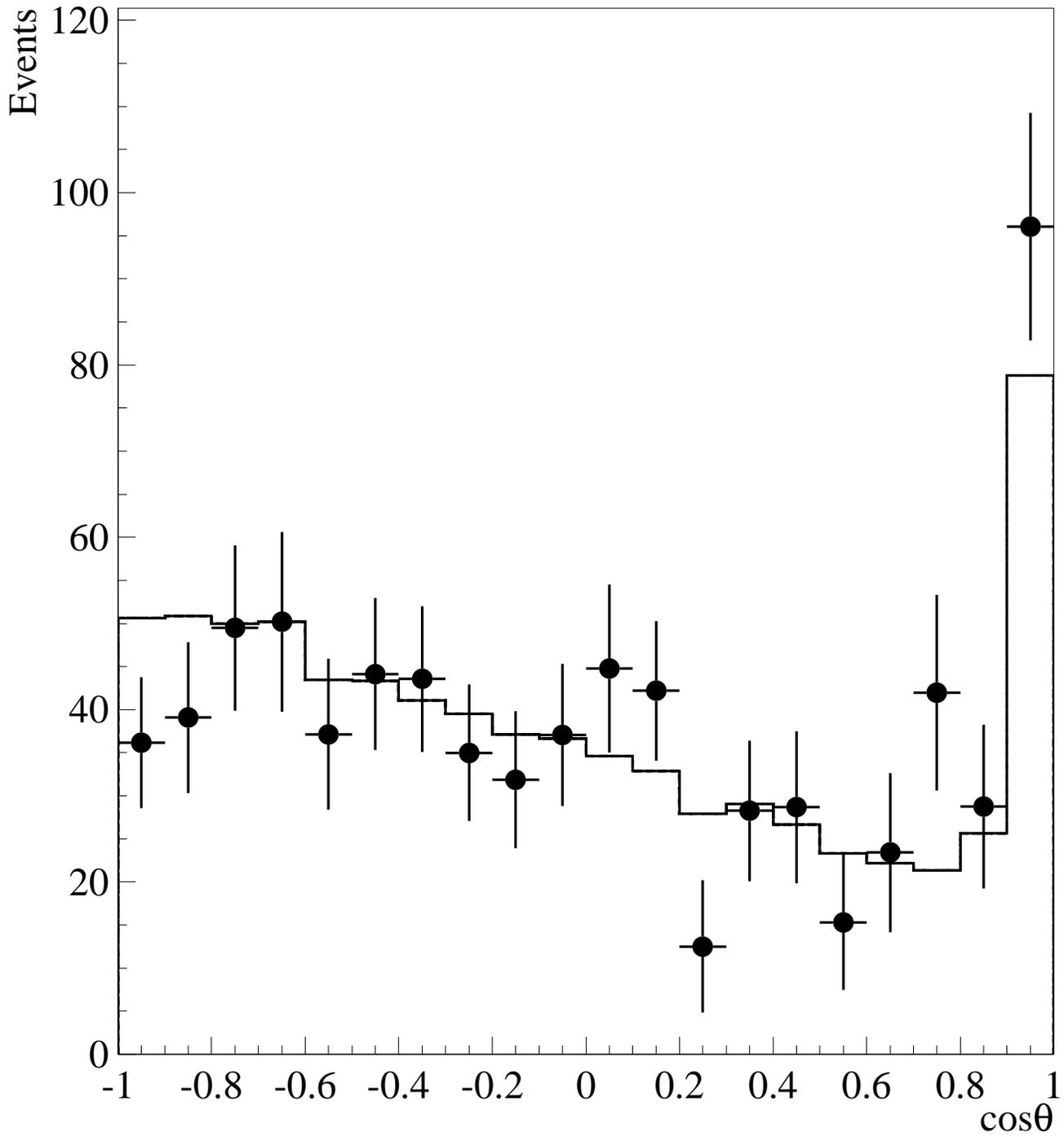,width=7.5in,silent=}}
\caption{The observed distribution of cos~$\theta$ compared 
with the expected (solid line) distribution.
Identified $^{12}\rm{N}_{g.s.}$ events are excluded.} 
\label{Fig. 20}
\end{figure}

\begin{figure}
\centerline{\psfig{figure=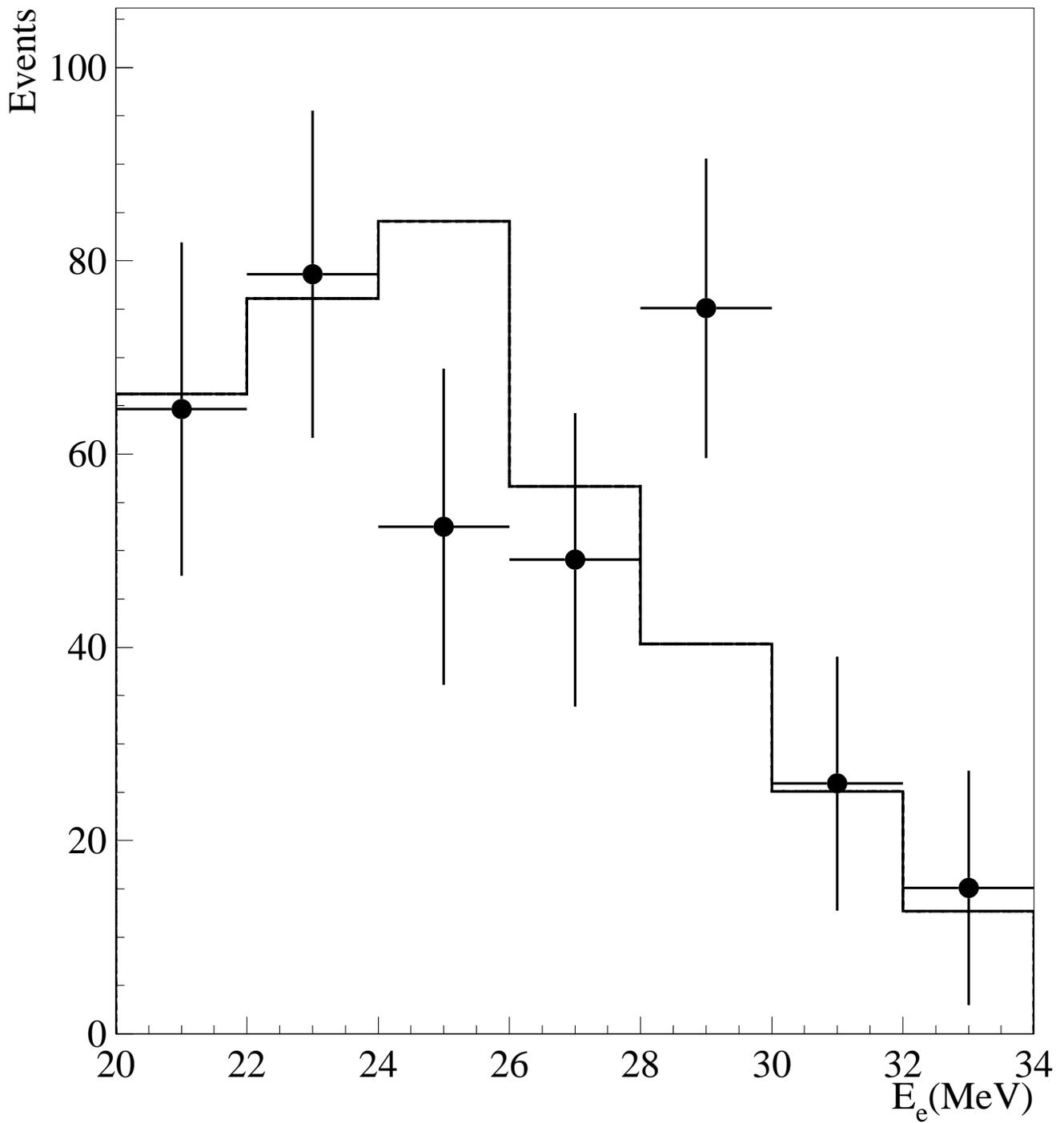,width=7.5in,silent=}}
\caption{The observed and expected (solid line) electron
energy distribution for the process
$^{12}\rm{C}(\nu_e, e^-)^{12}\rm{N}^*$. }
\label{Fig. 21}
\end{figure}

\begin{figure}
\centerline{\psfig{figure=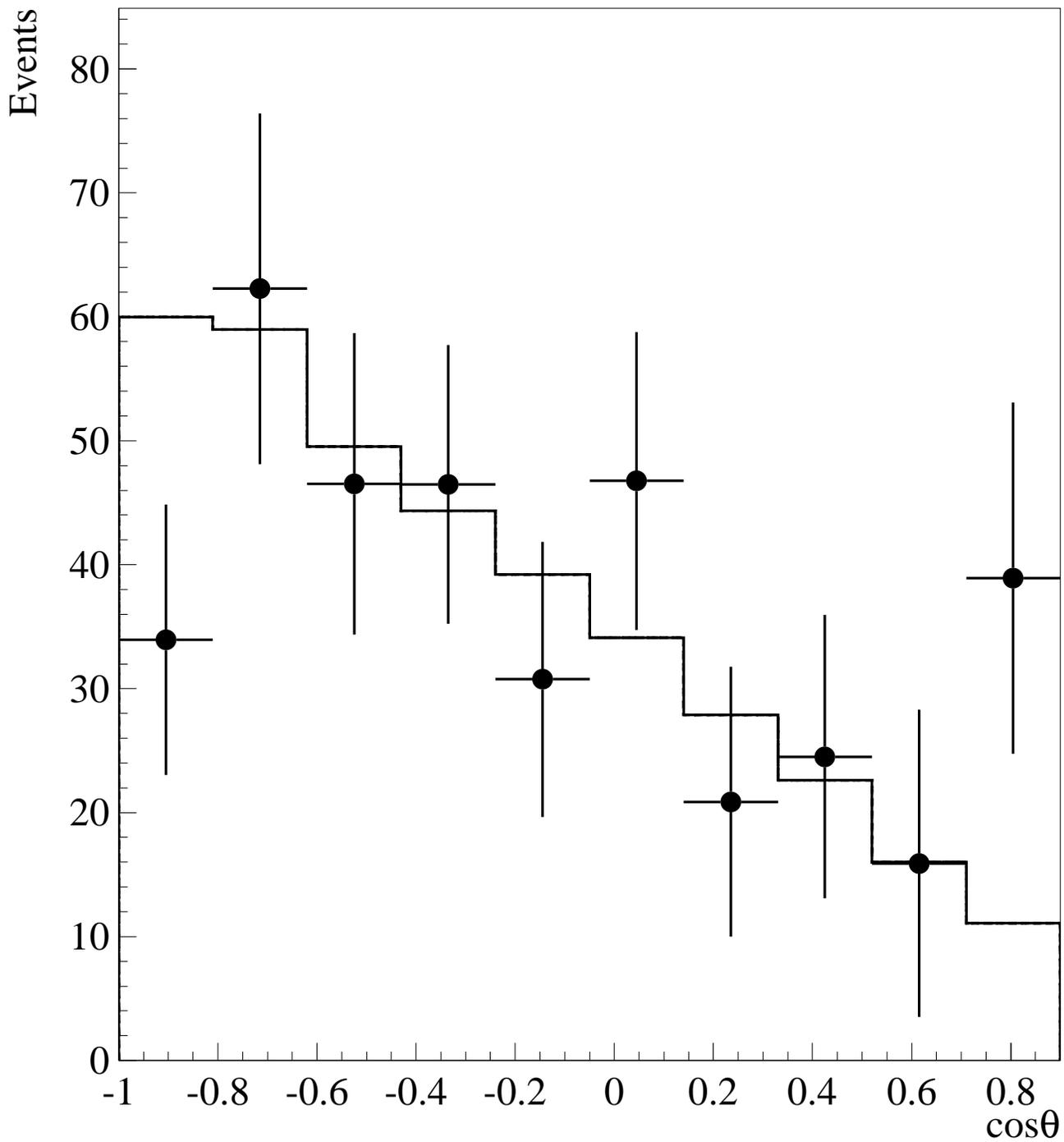,width=7.5in,silent=}}
\caption{The observed and expected (solid line) distribution of cos~$\theta$
for the process 
$^{12}\rm{C}(\nu_e, e^-)^{12}\rm{N}^*$. }
\label{Fig. 22}
\end{figure}

\begin{figure}
\centerline{\psfig{figure=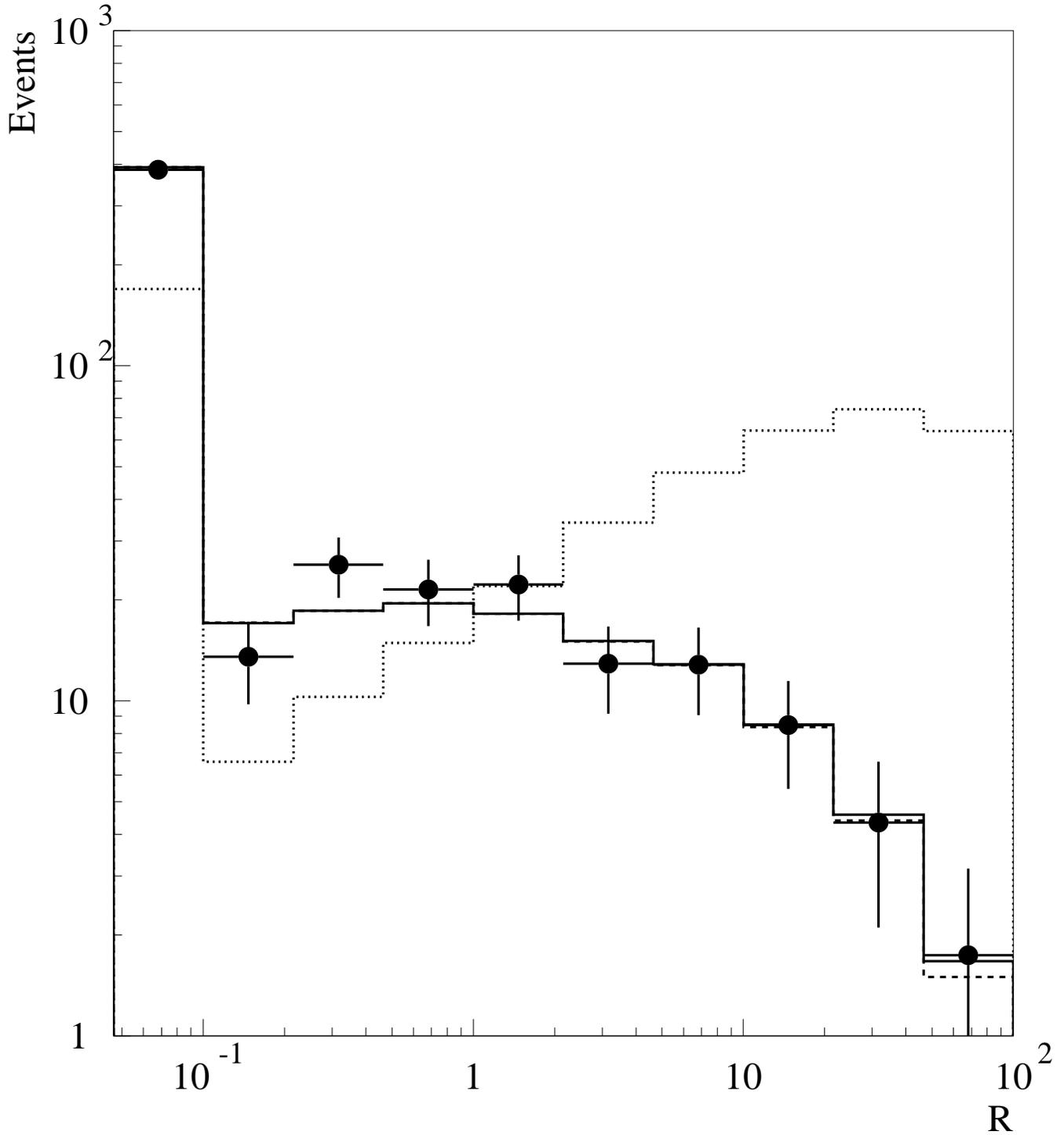,width=7.5in,silent=}}
\caption{The observed distribution of the $\gamma$ likelihood ratio R
 for the $^{12}\rm{C}(\nu_e, e^-)^{12}\rm{N}_{g.s.}$ sample.
 Shown for comparison are the  correlated distribution (dotted line), 
the uncorrelated distribution (dashed line) and
the best fit (solid line) to the data which has  a $(0.3 \pm 1.7)$\% correlated component.
}
\label{Fig. 23}
\end{figure}

\begin{figure}
\centerline{\psfig{figure=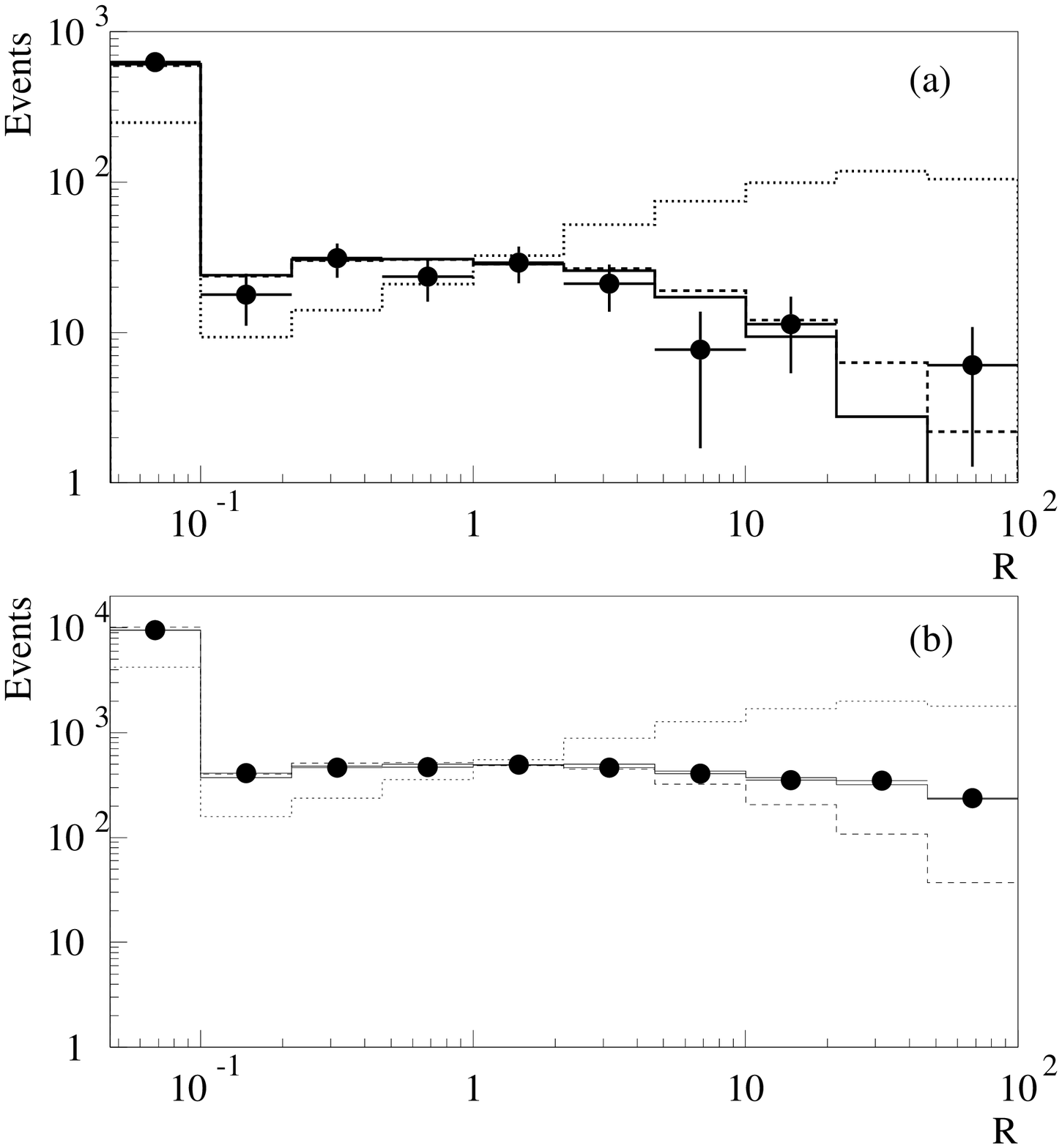,width=7.5in,silent=}}
\caption{The observed distribution of the $\gamma$ likelihood ratio R
 for the (a) beam-excess and (b) beam-off  sample of inclusive 
 electrons with cos~$\theta<0.9$. 
Identified  $^{12}\rm{N}_{g.s.}$ events are excluded.
 Shown for comparison are the  correlated distribution (dotted line), 
the uncorrelated distribution (dashed line) and
the best fit (solid line) to the data.
}
\label{Fig. 24}
\end{figure}

\begin{figure}
\centerline{\psfig{figure=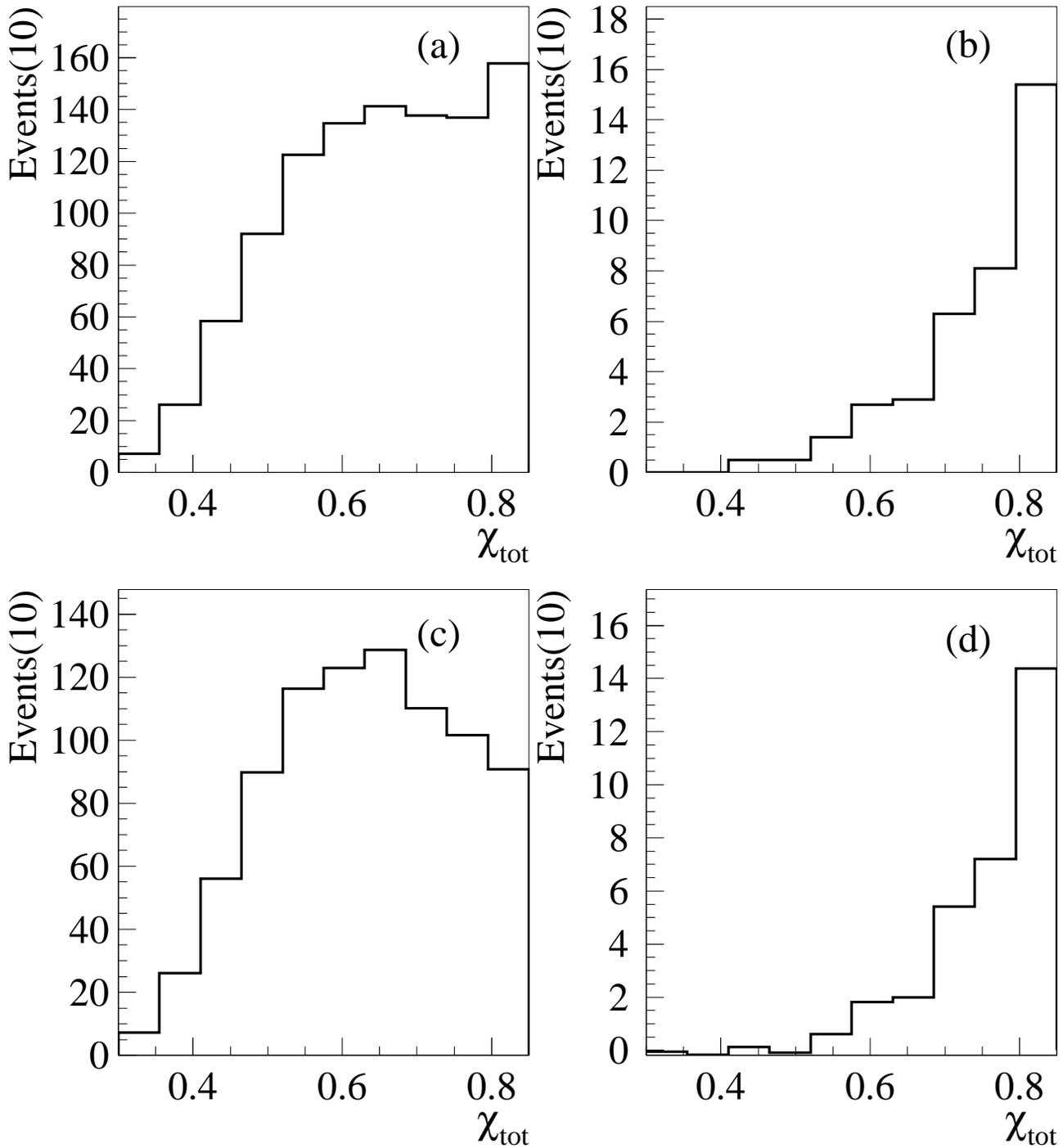,width=7.5in,silent=}}
\caption{Distribution of $\chi_{tot}$ for beam-off events
for (a) all events (b) events with R$>30$,
(c) uncorrelated component and (d) correlated component.}
\label{Fig. 25}
\end{figure}


\end{document}